\documentstyle[12pt,twoside]{article}
\input{amssym.def}
\input{amssym}
\newtheorem{theorem}{Theorem}
\newtheorem{proposition}{Proposition}
\newtheorem{lemma}{Lemma}

\def\LI{\Lambda_{\infty}}

\def\om{\omega}

\def\ha{Ha\-mil\-to\-nian}
\def\Sc{Schr\"o\-din\-ger}

\def\R{\Bbb R} 
\def\Z{\Bbb Z} 
 
\def\N{\Bbb N} 

\def\C{\Bbb C}
\def\Tr{{\rm Tr}}
\def\la{\langle}
\def\be{\begin{equation}}
\def\ee{\end{equation}}
\def\ra{\rangle}
\def\ds{\displaystyle}
\def\xxi{x,\xi}

\def\ah{\hat{a}}

\begin{document}
\date{}
%
%
%
%
\title{Ergodic Properties of
Infinite Harmonic \\ Crystals: an Analytic
Approach }

\author{
Sandro Graffi 
 \\Dipartimento di Matematica, 
Universit\`{a} di Bologna\\40127 Bologna, 
Italia
\\ {\tt\small graffi@dm.unibo.it}
 \and Andr\'e Martinez \\ 
 Universit\'{e} de Paris-Nord,  
D\'epartement de Math\'ematiques \\ CNRS -
URA 742, 93430 Villetaneuse, France
\\ {\tt\small martinez@math.paris-nord.fr}
\and {\tt\small} Expanded version of \\{\it An
Example of a Quantum Ergodic System}\\ {\tt\small
Bologna Mathematics Preprint 11-95, archived in
mp\_arc@math.utexas.edu 95-230}. }
\maketitle   
\begin{abstract}
 {We prove that the
quantum dynamics of a class of 
infinite harmonic crystals becomes ergodic
and mixing in the following sense: if $H_m$
is the $m$-particle Schr\"o\-dinger
operator, $\ds \omega_{\beta,m}(A)=
\Tr\,(A\exp{-\beta H_m})/\Tr\,(\exp{-\beta
H_m})$ the corresponding quantum Gibbs
distribution over the observables $A,B$,
$\psi_{m,\lambda}$ the coherent
states in the $m$-th particle Hilbert
space, $\ds g_{m,\lambda}=(\exp{-\beta H_m})
\psi_{m,\lambda}$
then   
$$
\lim_{t\to\infty}\lim_{n\to\infty}
\lim_{m\to\infty}
\frac1{T}\int_0^T\la e^{iH_nt}Ae^{-iH_nt}
\psi_{m,\lambda},\psi_{m,\lambda}\ra\,dt=
\lim_{m\to\infty}\omega_{\beta,m}(A) 
$$  
if the classical infinite dynamics is
ergodic, and 
$$
\lim_{t\to\infty}\lim_{n\to\infty}
\lim_{m\to\infty}
\omega_{\beta,m}(e^{iH_nt}Ae^{-iH_nt}B)=
\lim_{m\to\infty}\omega_{\beta,m}(A)\cdot
\lim_{m\to\infty}\omega_{\beta,m}(B) 
$$  
if it is in addition mixing. The
classical ergodicity and mixing properties
are recovered as $\hbar\to 0$, and
$\lim_{m\to\infty}\omega_{\beta,m}(A)$ turns
out to be the average over a classical Gibbs
measure of the symbol generating $A$ under
Weyl quantization. }   
\end{abstract}  
\baselineskip = 18pt 
\vfill\eject
\section{Introduction}
 \setcounter{equation}{0}%
\setcounter{theorem}{0}%
\setcounter{proposition}{0}%
\setcounter{lemma}{0}%
\setcounter{corollary}{0}%
\setcounter{definition}{0}%

This paper deals with the ergodic theory of a
 class of infinite 
quantum systems, the
harmonic crystals.
 In this introduction we review
the relevance of the infinitely 
many particle limit
in detecting chaotic behaviour
of quantum systems, state the results
and motivate  why to our opinion
is convenient to examine the problem
via pseudodifferential
operators. 
\par   Let
$H$ be the quantization of a \ha\ generating
a flow
$S_t$ on a constant energy manifold
$M_E\subset R^m$, $A, B\in{\cal L}({\cal H})$ any
suitable quantum observable in ${\cal
H}=L^2(\R^m)$, and let
$\sigma(H)$ be discrete and simple, with
projections $P_n$ on the eigenvectors $\{u_n:
n=0,1,\ldots\}$. If quantum chaotic behaviour
(if any) is to be characterized in terms of
ergodicity and mixing,
 we have to consider
the quantum microcanonical ensemble at energy
$E$, i.e. the application
$\omega_{\Delta,E}$ mapping any 
$A\in{\cal L}({\cal H})$ into:  
\begin{equation}
\label{qmc}
\omega_{\Delta,E}(A)=\frac{\Tr\, A\sum_{n:
E-\Delta<E_n<E}P_n}{\Tr\,\sum_{n:E-\Delta<E_n<E}P_n}
\equiv \frac{\Tr\,
A\delta(H-E)}{\Tr\, \delta(H-E)}
\end{equation}
(see
\cite{Ru} \S 1.3; $\Delta >0$ is
arbitrarily small). The quantum evolution ${\ds
A_H(t)=e^{iHt}Ae^{-iHt}}$ 
 of $A$  leaves
$\omega_{\Delta,E}(A)$ invariant. Hence the
 consequent definition of mixing is (see
Appendix 2 for details)      
\be
\label{qmcmixing}
\lim_{t\to\infty}\omega_{\Delta,E}(A_H(t)B)
=\omega_{\Delta,E}(A)\cdot\omega_{\Delta,E}(B)
\end{equation}
We can always find in ${\cal
H}$ (see Appendix 2 below for the easy
verification) a family of normalized vectors
$(\psi_{\lambda})_{\lambda\in{\Lambda}},
\Lambda=\R^{2m}$ complete for
$\omega_{\Delta,E}$, namely   
\be
\omega_{\Delta,E}(A)=\int_{\Lambda}\la
A\psi_{\lambda},\psi_{\lambda}\ra_{\cal
H}\,d\nu_{\Delta,E}(\lambda), \qquad
\forall\;A\in{\cal L}({\cal H})
\ee
for a well determined probability measure
$\nu_{\Delta,E}(\lambda)$ on $\Lambda$.
\par\noindent Then (\ref{qmcmixing})
becomes  
\be
\label{qmccmixing}
\int_\Lambda \la A_H(t)B\psi_\lambda ,
\psi_\lambda\ra_{\cal
H}d\nu_{\Delta,E}(\lambda ) \rightarrow
\int_\Lambda \la A\psi_\lambda ,
\psi_\lambda\ra_{\cal H}d\nu_{\Delta,E}(\lambda )
\int_\Lambda \la B\psi_\lambda ,
\psi_\lambda\ra_{\cal
H}d\nu_{\Delta,E}(\lambda) 
\ee
 as $\vert t\vert
\rightarrow \infty$.  This entails the 
following representation 
 of the quantum ergodicity notion 
(see again Appendix 2 ): for any 
$A\in{\cal L} ({\cal H})$ and for
$d\nu$-almost  all $\lambda\in\R^{2m}$,
\be
\label{qcerg}
\frac1T\int_0^T\la A_H(t)\psi_\lambda ,
\psi_\lambda\ra_{\cal H}dt \rightarrow 
\int_\Lambda \la A\psi_\lambda ,\psi_\lambda
\ra_{\cal H}d\nu_{\Delta,E} (\lambda
)=\omega_{\Delta,E}(A) 
\quad {\rm as}\quad
\vert T\vert \rightarrow \infty .
\ee 
On the other hand it is
well known (and easy to verify) that 
\be
\label{qtimeaverage}
\lim_{T\to\infty}\frac1{T}\int_0^T
\la\psi, A_H(t)\psi\ra\,dt=
\sum_{n=0}^{\infty}|\lambda_n|^2\la u_n,A u_n\ra
\ee
Here ${\ds
\psi=\sum_{n=0}^{\infty}\lambda_nu_n}$ is any
normalized quantum state expanded on the
eigenvector basis $(u_n)$. 
(\ref{qtimeaverage}) is the Von Neumannn
definition of quantum ergodicity\cite{VN} on the
microcanonical ensemble. 
Now the verification of (\ref{qcerg}) 
requires $H$ to have continuous spectrum
(\cite{Ru}, \S 1.3), and
(\ref{qtimeaverage}) shows
that the time average  cannot eliminate the
dependence on the initial datum 
${\ds
\psi\equiv\{\lambda_n\}_{n=0}^{\infty}}$. 
\par   
This {\it a priori} lacking of ergodicity, 
and {\it a fortiori} of mixing, looks
 as a manifestation of  the so called
"quantum suppression of classical chaos",
which however  can disappear when the number of
particles tends to infinity. This has 
been remarked in different contexts and within
different approaches in
\cite{Ch}, 
\cite{JLPC}, \cite{JL}, \cite{Be}. 
Hence the quantum counterparts of chaotic
systems with infinitely many degrees of 
freedom (for a recent review  see \cite{Be})
are the best candidates to look for
 chaotic behaviour. The simplest
 one is the 
infinite linear harmonic system  
\be 
\label{infinitesystem} \ddot{q}_i=
-2\sum_{i,j\in\Z}V_{ij}q_j \ee 
We prove that, when
the couplings $V_{ij}$ generate an infinite 
dimensional dynamics $\phi_t$
ergodic {\it with respect to the
(infinite  dimensional) Gibbs measure}
$d\mu_{G}(\beta)$ \cite{LL,Ti,VH}, the
 quantum evolution is ergodic, and
mixing if  $\phi_t$ is in
addition mixing. The
averages are now to be computed on the quantum
canonical ensemble (Gibbs
state at inverse temperature $\beta$),
i.e., the application $\omega_{\beta}$
mapping any  $A\in{\cal L}({\cal H})$
into $\ds 
\omega_{\beta}(A)=\frac{\Tr\,ÊA e^{-\beta H}}
{\Tr\, e^{-\beta H}}  .\;$
More precisely, denote:   
\be
\label{q_m} 
q_m(\xxi)=\frac12|\xi|^2+\la V_m x,x\ra, 
\qquad V_m=
\left(V_{i,j}\right)_{{|i|\leq m}
\atop {|j|\leq m}}
\ee
the $(2m+1)$ dimensional \ha\ defined on
$\Lambda_m=(\R^{2m+1})^2$;  
$H_m=Op^W(q_m)$ the operator on  
$L^2_m\equiv L^2(\R^{2m+1})$
defined by its Weyl quantization, $A=Op^W(a)$
the   operator on
$L^2_m$ quantizing $a\circ\Pi_{m_1}(\xxi)$
($m_1$ fixed) where $a$ is any smooth
classical observable
on
$\Lambda_{m_1}$ and  $\Pi_{m_1}(\xxi)\equiv
(\xxi)_{|i|\leq m_1}$. \par Then the
present
 results are
 (see Theorems 2.2, 2.3  and Proposition
5.1 for a sharper version): $\forall \beta
>0$ 
\be
\label{1.5} 
\lim_{T\to\infty}\lim_{n\to\infty}\lim_{m\to\infty}
\frac1{T}\int_0^T\left\la
A_{n}(t)\psi_{\lambda,m},
 \psi_{\lambda,m}\right\ra
_{L^2_m}dt
 = 
\lim_{m\to\infty}\int_{\Lambda_{\infty}}\la
A \psi_{\lambda,m},
\psi_{\lambda,m}\ra_{L^2_m}
 d{\nu}_{m}(\lambda)
\ee
for ${\nu}$-almost any $\lambda$, 
and
\be
\label{1.6}
\lim_{t\to\infty}\lim_{n\to\infty}\lim_{m\to\infty}
\omega_{\beta,m}(A_n(t)B)= \lim_{m\to\infty}
\omega_{\beta,m}(A)\cdot\lim_{m\to\infty}
\omega_{\beta,m}(B)
\ee
Here:  
\be
\label{aam}
\omega_{\beta,m}(A) = \frac{\Tr\,
A e^{-\beta H_m}} {\Tr\, e^{-\beta
H_m}};\quad
\omega_{\beta,m}(A_n(t)B)=
\frac{\Tr\, A_{n}(t)B e^{-\beta
H_m}} {\Tr\, e^{-\beta H_m}};
\ee
$A_{n}(t)$ is the Heisenberg
observable corresponding to $A$ under the
quantum evolution of
$H_n$;
\par\noindent
$$
 \psi_{\lambda,m}=
\frac{\exp ({-\beta H_m/2})f_{\lambda,m}}
{\|\exp ({-\beta
H_m/2})f_{\lambda,m}\|}\quad
\nu_{m}(\lambda)=
\frac{\|\exp ({-\beta
H_m/2})f_{\lambda,m}\|^2
\,d\lambda}
{\int_{\Lambda_m}\|\exp ({-\beta H_m/2})
f_{\lambda,m}\|^2\,d\lambda} 
$$ 
$f_{\lambda,m}$ being the Bargmann coherent
states (a set of vectors in
$L^2(\R^{2m+1})$ indexed by
$\lambda\in\Lambda_m$ whose definition is
recalled in Appendix 2);
$\nu(\lambda)=\lim_{m\to\infty}\nu_m(\lambda)$.
\vskip 0.2cm\noindent
 {\bf Remark 1.} 
 The mixing property with respect 
to the KMS states in the CCR algebra of the
infinite harmonic crystal (which has the
same $W^*$ closure of the pseudifferential
algebra we use) is proved in
\cite{Be}, Example 4.46, through
the asymptotic abelianess of the
Weyl algebra automorphism generated by the
dynamics of the infinite crystal, when 
$\sigma (V)$
is purely absolutely continuous so
that classical mixing holds
\cite{LL}. The asymptotic abelianess may
however
 fail if
$\sigma (V)$ is only continuous and the
classical system is only ergodic. Hence
the ergodicity result (\ref{1.5}) requires in
general an independent proof. 
\vskip 0.1cm\noindent
{\bf Remark 2.} The main reason  why, to our
opinion,  an "analytic" proof, based on
pseudodifferential calculus, is in any case
useful is that the notion
(\ref{qmccmixing}) is proved to have the
expected classical limit (Appendix 2). 
\vskip 0.1cm\noindent
Additional
reasons are the following: 
\begin{enumerate}    
\item  One finds the r.h.sides
of (\ref{1.5}),(\ref{1.6})  to be the
relevant classical averages: 
\be
\lim_{m\to\infty}\int_{\Lambda_{\infty}}\la
A_{m_1} \psi_{\lambda,m},
\psi_{\lambda,m}\ra\,
d{\nu}_{m}(\lambda)=
\int_{\Lambda_{\infty}}a\circ\Pi_{m_1}
\,d\hat\mu_{\beta}
\ee
\be
   \lim_{m\to\infty}\frac{\Tr\, A_{m_1}e^{-\beta
H_m}} {\Tr\, e^{-\beta H_m}}\cdot
\frac{\Tr\,B_{m_1}e^{-\beta H_m}}
{\Tr\, e^{-\beta H_m}}=
\int_{\Lambda_{\infty}}a\circ\Pi_{m_1}
\,d\hat\mu_{\beta}\cdot
\int_{\Lambda_{\infty}}b\circ\Pi_{m_1}
\,d\hat\mu_{\beta}
\ee
Here
$\hat\mu_{\beta}=\lim_{m\to\infty}\hat\mu_{\beta,m}$,
where $\hat\mu_{\beta,m}$ is the (explicitly
constructed) Gibbs measure on $\Lambda$ whose
Weyl quantization yields $e^{-\beta H_m}$.
It turns out that $\hat\mu_{\beta}$
depends on $\hbar$ and reduces to
$\mu_{G}(\beta)$ as
$\hbar\to 0$, because
$e^{-\beta q_m}$ is just the principal symbol of
$e^{-\beta H_m}$ realized as a pseudodifferential
operator.
\item
If the initial states $\psi_{m,\lambda}$
belong to an explicitly constructed set (the
image under  $e^{-\beta H_m}$ of
"almost all" coherent
states on
$\Lambda_m$), the $m\to\infty$ limit can
actually eliminate the dependence of the
r.h.s. (\ref{1.5}) on the particular
state in the set.  
\item
Unlike the algebraic proof, the analytic one
can be  in principle extended to systems
quantizing non-linear classical
equations. Work in this direction is in
progress: it can be proved
\cite{GJLM} that in some non linear cases
the above results are still true in the
sense of the formal power series in
$\hbar$.  
\end{enumerate}   
 We conclude this introduction with  the
remark that the dynamical mechanism generating
chaotic  behaviour, in the classical case
and in the quantum one as well, is but free
propagation of the chaotic initial condition: 
the infinite harmonic crystal goes indeed over
(when the spacing goes to zero, and for
special choices of $V$) to the free
 wave equation (equivalently, there exist
coordinates in which the particle 
motions are free)
and the chaotic initial condition is
selected by the invariant
 Gibbs measure. 
This situation is referred to as kinematic
chaos \cite{JLP2}. \par The paper is
organized as follows: in the next Section we
state assumptions and results, after a brief
recall of the infinite dimensional classical
harmonic dynamics; in \S 3 and in \S 4 we prove
the quantum ergodicity and the quantum
mixing, respectively, in the most general
formulation. In \S 5 and \S 6 we prove
a sharper formulation of the above results
when $\exp{-\beta H_m}$ is replaced by
$Op^W(\exp{-\beta q_m})$ and the
family of vectors in $L^2_m$ is specialized to
the coherent states.  Appendix 1 contains the
proof of some technical lemmas, and
Appendix 2 contains the discussion of our
results in the light of the existing notions
of quantum ergodicity and mixing, together
with the verification that they have the
expected classical limit.  
\vskip 0.2cm\noindent
{\bf Acknowledgments.} We thank
 G.Jona Lasinio for many illuminating
discussions, and an anonymous referee for
pointing  out to us the relevant results out
of the $W^*$ dynamical systems.

\section{
Assumptions and Statement of the 
Results}    
\setcounter{equation}{0}%
\setcounter{theorem}{0}%
\setcounter{proposition}{0}%
\setcounter{lemma}{0}%
\setcounter{corollary}{0}%
\setcounter{definition}{0}%
 
In the notation
of
\cite{LL}, to which we refer the reader for
any further detail on the system of
infinitely many oscillators,  let ${\ds
V=\left(V_{i,j}\right)_{i,j\in \Z}}$ be an 
infinite real-symmetric matrix;  $q_m$ and
$V_m$ are as in (\ref{q_m}) and 
$\Lambda_m=(\R^{2m+1})^2$. \par We write
$S_m(1)$ for the set of  $C^\infty$
functions on $\Lambda_m$ which are bounded
together with all their derivatives, and for
$a\in S_m(1)$ we denote $Op^W(a)$ the Weyl
quantization  (with $\hbar=1$) of the symbol
(equivalently, classical observable) $a$,
explicitly given by the oscillatory integral
:  
\be 
\label{Weyl} Op^W(a)u(x)
=(2\pi)^{-(2m+1)}\int_{\Lambda_m}e^{i\la
(x-y),\xi\ra}a(\frac{x+y}{2},\xi)u(y)\,dy\,d\xi
\ee for all $u\in {\cal S}(\R^{2m+1})$.
In particular the \Sc\ operator $H_m$   on
$L^2(\R^{2m+1})$
\be
\label{H_m} H_m := Op^W(q_m)=
\frac{1}{2}(\sum_{j=1}^{2m+1}
 D_{x_j}^2)+\la V_m x,x\ra, \qquad
D_{x_j}=-i\frac{\partial}{\partial x_j}
\ee
quantizes the  \ha\ 
$q_m$ describing $m$ oscillators coupled
through $V_m$.\par We assume from now on
\begin{itemize}
\item [(H1)]
\hskip 0.9cm
$\qquad
|V_{ij}|={\cal O}(|i-j|^{-\infty}), 
\qquad |i-j|\to +\infty
$ \hfill\break\vskip 0.1cm 
and $\exists \, 0 <\varepsilon
<M<\infty$  such that $\forall m\geq 0$,
$\sigma(V_m)\subset [\varepsilon ,M]$.
\end{itemize} 
In particular, $V:\ell^2(\Z)\rightarrow 
\ell^2(\Z)$
is bounded and strictly  positive, with
  $\sigma(V)\subset[\varepsilon ,M]$. 
\begin{itemize} \item [(H2)] The operator
$V$ acting on $\ell^2(\Z)$ has  no point
spectrum. \end{itemize} Denote %
\be
\label{Lambda_infinity}
\Lambda_{\infty}:=\bigcup_{k\in\N}
\left\{(x_j,\xi_j)\};\qquad |x_j|+|\xi_j|={\cal
O}(|j|^k), \; 
|j|\to\infty\right\}:=\bigcup_{k\in\N}{\cal H}_k
\ee
It is proved in \cite{LL} that under condition
(H1), 
$\Lambda_{\infty}$ is invariant under the classical
evolution  of infinitely many degrees of freedom
defined as follows
\be
\label{Phi_t}
\phi_t(\xxi)\equiv \phi(t,\xxi)=e^{tB}(\xxi), 
\qquad \forall (\xxi)\in\Lambda_{\infty}, \forall
t\in\R
\ee
where $B(\xxi)$ is the infinite-dimensional 
\ha\ vector field generated by
$q_m$ when $m\to\infty$
\be
\label{A(xxi)}  B(\xxi)=\left(\xi_j,-2\sum_{k\in\Z}
V_{jk}x_k\right)_{j\in\Z}
\ee
 Moreover, if
$\Pi_m:\Lambda_{\infty}\to\Lambda_m$  denotes the
projection
\be
\label{Pi_m}
\Pi_m(\xxi)=(x_j,\xi_j)_{|j|\leq m}
\ee
for any $(\xxi)\in{\cal H}_k$ one has
\be
\label{1}
\phi(t,\xxi)=\lim_{m\to\infty}\phi_{m,t}(\Pi_m(\xxi))\in{\cal
H}_k
\ee
where $\phi_{m,t}={\rm \exp} tH_{q_m}$, ${\ds
H_{q_m}=\left(\frac{\partial q_m} {\partial
\xi},-\frac{\partial q_m}{\partial x}\right)}$ is
the  vector  field generated by $q_m$, and
the limit is taken with respect to the
natural Banach space topology  of ${\cal
H}_k$.\par  Now by (H1) the operator
$V^{-\frac12}$ exists and  is continuous on
$\ell^2(\Z)$. This assumption and (H2)  allow
Lanford and Lebowitz\cite{LL} to prove the
existence of the infinite dimensional,
 ergodic Gibbs measure
$d\mu_G(\beta)$ on $\LI$, namely
\begin{enumerate}
\item
\be
\label{2}
\int_{\LI}\varphi\circ\Pi_{m_1}\, d\mu_G(\beta)=
\lim_{m\to\infty}\int_{\Lambda_m}
\varphi\circ\Pi_{m_1} (\xxi)e^{-\beta q_m(\xxi)}\,
\frac{dx\,d\xi}{Z_m}, 
\quad \forall\varphi\in C^0_b(\R^{2m_1+1})
\ee
 where
\be
\label{Z_m} Z_m(\beta)=\int_{\Lambda_m} e^{-\beta
q_m(\xxi)}\,dx\,d\xi
\ee
is the $m$-particle partition function;
\item The Gibbs measure is invariant and ergodic 
with respect to the flow
$\phi(t;\xxi)$, namely the continuous dynamical 
system
$(\LI,\phi_t,d\mu_G(\beta))$ is ergodic.
\end{enumerate}
An example of an infinite matrix satisfying 
(H1)-(H2) is given by $V=W$ where
\be
\label{ex} W_{ij}=0, \qquad |i-j|\geq 2, \qquad
W_{ii}=1, 
\qquad W_{i,i+1}=W_{i,i-1}=\alpha
\ee
with ${\ds |\alpha|<\frac12}$, $\alpha\in\R$. 
The properties (H1), (H2) are proved e.g. in
\cite{Sj}.
\vskip 0.2cm\noindent
\par To state our result we need to establish  some
further notation. For $f\in L^2(\R^{2m+1})$ and
$(\xxi)\in\Lambda_m$,  we introduce the Wigner
function of $f$
\be
\label{w_f}
w_f(\xxi)=\int_{\R^{2m+1}}e^{iu\xi}f(x-\frac{u}{2})
\overline{f(x+\frac{u}{2})}\,du
\ee
and we restrict our attention to a random  set
of states $f$ in the following sense :
for all $m\in\N$, we consider a measure space 
$(X_m,\theta_m)$ with positive measure
$\theta_m$, and a family
$(f_\lambda)_{\lambda\in X_m}$   of functions in
$L^2(\R^{2m+1})$ such that :
\begin{itemize}
\item[(H3)] For $dxd\xi$-almost all 
$(\xxi )\in\Lambda_m$, the application
$X_m\ni\lambda\mapsto w_{f_\lambda}(\xxi )$ is  in
$L^1(X_m , d\theta_m  )$ with non negative values,
and the quantity
$\ds\int_{X_m} w_{f_\lambda}(\xxi ) d\theta_m
(\lambda )$ is  ($dxd\xi$-almost everywhere)
constant with respect to $(\xxi )$.
\end{itemize} Here we can notice that, at least
formally,  (H3) is implied by the property (to be
compared with (\ref{3'})):
$${\rm Tr}(A)= \int \la Af_\lambda ,f_\lambda\ra
d\theta_m (\lambda )$$ for any trace-class operator
$A$. Indeed we have $w_{f_\lambda}(\xxi )=
\la A_{\xxi} f_\lambda ,f_\lambda\ra$ with $A_{\xxi}
f(y)=e^{2i(y-x)\xi}f(2x-y)$, which actually is not
trace-class, but whose distributional kernel $\ds
K_{\xxi} (y,y')= e^{2i(y-x)\xi}\delta (y'+y=2x)$
formally satisfies:
$\ds\int K_{\xxi} (y,y)dy=1$.

 In the last section we develop an example  (the
so-called coherent states) where (H3) is satisfied.
Note that in any case, 
$w_f(\xxi )$ is real and satisfies :
$\ds\int w_f(\xxi )dxd\xi  =(2\pi)^{2m+1}\| f\|^2$. 
\vskip 0.5cm As we shall see, (H3) implies among
other things that
\be
\label{fini}
\int_{X_m}\| e^{-\beta H_m/2}
f_\lambda\|^2d\theta_m (\lambda )< +\infty
\ee
so that we can consider the following
probability measure on $X_m$:
\be
\label{nutilde} d\nu_m(\lambda )=\frac{\|e^{-\beta
H_m/2} f_\lambda\|^2d\theta_m (\lambda )}{\int_{X_m}
\|e^{-\beta H_m/2}f_\lambda\|^2d\theta_m  (\lambda
)}.
\ee
\par Now let 
\be W_{\beta}=\sqrt{2}V^{-\frac12}{\rm tanh}
\frac{\beta V^{\frac12}}{\sqrt{2}}
\ee
(which is well defined on $\ell^2(\Z )$),  and
for $\beta >0$, denote $\hat\mu_{\beta}$ the
Gaussian probability measure on $\Lambda_\infty$ 
with mean zero and covariance given by:
\begin{eqnarray}
\label{muhat} &&E[x_ix_j]=\langle (2VW_\beta
)^{-1}e_i , e_j\rangle_{\ell^2(\Z)}\nonumber\\ &&
E[\xi_i \xi_j]=\langle W_\beta^{-1}e_i ,
e_j\rangle_{\ell^2(\Z)}\\ && E[x_i\xi_j]=0\nonumber
\end{eqnarray}
where $e_i=(\delta_{ij})_{j\in\Z}$.  Then our
first main result is :
\begin{theorem}
\label{mainth} Assume {\rm (H1)-(H3)}. Then :
\par (i) For any $\beta >0$, the dynamical system 
$(\Lambda_\infty , \phi_t, \hat\mu_\beta )$ is
ergodic ;
\par (ii) For $m_1\in\N$ fixed and $a\in
S_{m_1}(1)$, denote
$$ g_{m,\beta ,\lambda}=
\frac{e^{-\frac12\beta H_m}f_\lambda}{\|  e^{-
\frac12 \beta H_m}f_\lambda\|}
$$  
and
$$ A(m,n,T,\lambda )=\frac1{T}\int_0^T\left\la
e^{itH_n}Op^W(a\circ\Pi_{m_1})e^{-itH_n} 
g_{m,\beta ,\lambda}\, ,\, g_{m,\beta
,\lambda}\right\ra _{L^2(\R^{2m+1})}\,dt. 
$$ 
Then one has
$$
\lim_{T\to\infty}\limsup_{n\to\infty}
\limsup_{m\to\infty}\int_{X_m}\left\vert
A(m,n,T,\lambda )-\int_{\Lambda_\infty} a\circ
\Pi_{m_1}\, d\hat\mu_{\beta}\right\vert
d\nu_m(\lambda )=0.
$$ 
\end{theorem}
\vskip 0.5cm {\bf Remarks.}
\begin{enumerate}
\item  $A(m,n,T,\lambda )$ can be
made  arbitrarily close to $\ds \int a\circ
\Pi_{m_1}\, d\hat\mu_{\beta}$ in 
$L^1(X_m,d\nu_m(\lambda ))$ by first choosing 
$T$, then $n=n(T)$, and finally $m=m(n,T)$
 large enough. The {\it
pointwise} convergence of $A(m,n,T,\lambda
)$ is proved in Proposition  5.1 below,
choosing for $f_\lambda$ a particular set
of coherent states. %
\item Note that $A(m,n,T,\lambda )$ is well
defined   since the action of
$e^{itH_n}Op^W(a\circ\Pi_{m_1})e^{-itH_n}$ on 
$g_{m,\beta ,\lambda}$ which is a $C^\infty$
function on
$\R^{2m+1}$ is well defined for $n\leq m$.
\item For $\beta$ small, we have $W_\beta = \beta I
+ {\cal O}(\beta^3)$ and therefore the covariance
of $\hat\mu_\beta$ coincides with the one of the 
usual Gibbs measure $\mu_G(\beta )$ up to a
${\cal O}(\beta^3)$-error term. In this sense,  we
can say that $\hat\mu_\beta$ and $\mu_G(\beta )$
are asymptotically equal for small $\beta$'s (that 
is for large temperatures).
\item The measure $\hat\mu_\beta$ can be seen as
the limit
 when $m\rightarrow +\infty$ of the probability
measure on $\Lambda_m$ obtained by normalizing  
$\displaystyle e^{-q_{\beta ,m}(\xxi )}dxd\xi$,
where
\be q_{\beta ,m}(\xxi )=q_m(W_{\beta ,m}^{1/2}x,
W_{\beta ,m}^{1/2}
\xi )
\ee
and
\be 
W_{\beta ,m}=\sqrt{2}V_m^{-\frac12}{\rm tanh}
\frac{\beta V_m^{\frac12}}{\sqrt{2}}.
\ee
(In fact, one can prove that (H1) and the spectral
theorem imply that for any continuous function $f$
on
$\R$, $\la f(V_m)e_i,e_j\ra$ tends to  $\la
f(V)e_i,e_j\ra$ as $m\rightarrow\infty$.)

This relation between
$\hat\mu_\beta$ and the usual  Gibbs measure
reflects the relation between the usual quantum
Gibbs measure 
$e^{-\beta H_m}$ and the Weyl quantization of the
classical Gibbs measure
$e^{-\beta q_m}$, namely (see Lemma 
\ref{lemma} below):
$$
 e^{-\beta H_m}=C_{\beta,m}Op^W(e^{-q_{\beta ,m}})
$$
where $C_{\beta ,m}$ is a constant. In
particular, if
 we denote $\#$ the Weyl composition of symbols, we
get (with some other constant 
$C_{\beta ,m}'$):
\be
\label{2beta} e^{-q_{\beta ,m}}\# e^{-q_{\beta ,m}}
=C_{\beta ,m}' e^{-q_{2\beta ,m}}
\ee
which also explains the fact that
$\hat\mu_{\beta}$ appears in the result given the
above choice of
$g_{m,\beta,\lambda}$, dictated by the standard
requirement $\ds \Tr\,e^{-\beta H_m}<+\infty$. 
\end{enumerate}
To state the mixing property we need  two
additional  assumptions
\begin{itemize}
\item[(H4)] The spectrum of $V$ on $\ell^2(\Z )$ is
absolutely continuous.
\end{itemize}
\begin{itemize}
\item[(H5)] The matrix
$\ds W_{\beta ,m}=\sqrt{2}V_m^{-\frac12}{\rm tanh}
\frac{\beta V_m^{\frac12}}{\sqrt{2}}$ satisfies:
$$
(W_{\beta ,m})_{i,j}={\cal O}(|i-j|^{-\infty}) 
$$ 
uniformly with respect to $m$, $i$ and $j$.
\end{itemize} Note that (H5) is satisfied e.g. for
$V$ of the form $V=I+\alpha J$ where $J$ admits only
a finite number of non-zero diagonals and
$\alpha\in\R$ is chosen small enough. In
particular, the example given in (\ref{ex})
satisfies (H1) and  (H4)-(H5) if
$\vert\alpha\vert$ is small enough. Note also that
the absolute continuity of
$\sigma(V)$ implies  (\cite{LL}) that the
continuous dynamical system $(\LI,\phi_t,\mu_G)$ 
enjoys the mixing property.
For $m_1\in\N$ and $a\in
 S_{m_1}(1)$, we denote $\hat a\in {\cal
S}'(\Lambda_{m_1})$ the usual Fourier transform of
$a$ formally given by the integral:
\be
\hat a(x^*,\xi^*)=\int_{
\Lambda_{m_1}}e^{-i\la (x,\xi ),
(x^*,\xi^*)\ra}a(x,\xi ) dxd\xi .
\ee Then the result is:
\begin{theorem}
\label{th2}  Assume {\rm (H1)} and {\rm
(H4)-(H5)}. For $m_1\in\N$ fixed, $a,b\in
 S_{m_1}(1)$, and $n\geq m_1$  denote
\begin{eqnarray*} && A = Op^W(a\circ\Pi_{m_1})\\ &&
A_{n}(t)= e^{itH_n}Ae^{-itH_n} =
Op^W(a\circ\Pi_{m_1}\circ\phi_{n,t}) :=
Op^W(a_{n,t}) \\ && B=Op^W(b\circ\Pi_{m_1}).
\end{eqnarray*}
 Then we have
\be
\label{th3.2a}
\lim_{m\to\infty}\frac{\Tr (Ae^{-\beta H_m})}{\Tr
(e^{-\beta H_m})}=\int_{\Lambda_\infty} a\circ
\Pi_{m_1}\, d\hat\mu_{\beta}:=\omega_\beta (A)
\ee
and if moreover $\hat a$ and $\hat b$ are bounded
measures on $\Lambda_{m_1}$, one has:
\be
\label{rmixing}
\lim_{t\to\infty}\lim_{n\to\infty}
\omega_\beta \big( A_n(t)B\big) = \omega_\beta (
A)\cdot
\omega_\beta ( B)
\ee
\end{theorem}
\noindent
{\bf Remark.} Although this corresponds to the
notion of quantum mixing already existing in the
framework of
$W^*$ dynamical systems, our procedure permits us
to completely avoid to realize any algebra of
operators on an infinite dimensional space.

\vskip 0.5cm

\par Under an additional assumption on $V$ the
results
 of Theorems 2.1 and 2.2 admit a less cumbersome
formulation which eliminates the necessity of  the
double limit with respect to $m$ and $n$. The
further assumption is :
\begin{itemize}
\item[(H6)] For all $m\geq 0$ there exists a 
$(2m+1)\times (2m+1)$ real-symmetric matrix
$\tilde V_m$ satisfying the same assumption  (H1)
as $V_m$, and such that :\hfill\break (i) $\forall
i,j\in\Z$, 
$\langle {\tilde V}_m^{-1}e_i,e_j\rangle$ tends to 
$\langle { V}^{-1}e_i,e_j\rangle_{\ell^2}$ as 
$m\rightarrow +\infty$ ;\hfill\break (ii) $\forall
m,n\in\Z$, the operator 
$\Pi_m{\tilde V}_n^{\frac12}\Pi_n$ becomes
independent of $n$ for $n$ sufficiently large
;\hfill\break  (iii) The operator ${\tilde
V}_m^{-\frac12} \Pi_m{\tilde V}_n^{\frac12}\Pi_n$
tends strongly to the identity  on each ${\cal
H}_k$ ($k\in \N$) as $m\rightarrow +\infty$.
\end{itemize}
\par It is not very difficult to verify that an 
example of such $V$ satisfying (H4) (in addition to
(H1)-(H2)) is given by $V=W^2$ where $W$  is as in
the example (\ref{ex}) : in this case one can take
$\tilde V_m = (W_m)^2$ where $W_m$  is extracted
from $W$ as in (\ref{q_m}).
\vskip 0.5cm
\par Under assumption (H6), we define for 
$0<\beta <(2M)^{-1/2}$ :
\be
\label{qtilde}
\tilde q_{\beta ,m}(x, \xi )=\langle F_{\beta ,m}
\xi ,\xi\rangle +\langle G_{\beta ,m}x ,x\rangle
\ee
where (denoting $I_m$ the identity on
$\R^{2m+1}$):
\begin{eqnarray}
\label{F} && F_{\beta ,m} = \frac1{2\beta}
\tilde V_m^{-1}\left( I_m - (I_m-2\beta^2\tilde
V_m)^{\frac12}
\right)\\
\label{G} && G_{\beta ,m} = 2\tilde V_mF_{\beta ,m}.
\end{eqnarray}
We also consider on $X_m$ the probability measure:
\be
\label{nutilde'}
 d\tilde\nu_m(\lambda )=\frac{\|Op^W (e^{-\tilde
q_{\beta ,m}})f_\lambda\|^2 d\theta_m (\lambda
)}{\int_{X_m}
\|Op^W(e^{-\tilde q_{\beta ,m}})
f_\lambda\|^2d\theta_m (\lambda )}.
\ee
Then the result is:
\begin{theorem}
\label{th3} Assume {\rm (H1)-(H3)} and {\rm (H6)},
and, for $m_1\in\N$ fixed, $a\in
 S_{m_1}(1)$ and $0<\beta <(2M)^{-1/2}$, denote
$$
\tilde g_{m,\beta ,\lambda}=\frac {Op^W(e^{-\tilde
q_{\beta /2 ,m}})f_\lambda}{\| Op^W(e^{-\tilde
q_{\beta /2 ,m}})f_\lambda\|}
$$  
and
$$
\tilde A(m,T,\lambda )=\frac1{T}\int_0^T\left\la
e^{itH_m}Op^W(a\circ\Pi_{m_1})e^{-itH_m}  
\tilde g_{m,\beta ,\lambda}\, ,\,
\tilde g_{m,\beta ,\lambda}
\right\ra _{L^2(\R^{2m+1})}\,dt .
$$ 
Then one has
\begin{itemize}
\item[(i)]
$$
\lim_{T\to\infty}\limsup_{m\to\infty}
\int_{X_m}\left\vert
\tilde A(m,T,\lambda )-\int_{\Lambda_\infty}  a\circ
\Pi_{m_1}\, d\mu_G({\beta})\right\vert
d\tilde\nu_m(\lambda )=0.
$$ 
\item[(iii)] Assume furthermore (H4). Then
(\ref{th3.2a}) becomes
\be
\label{th3.3a}
\lim_{m\to\infty}\frac{\Tr A\,Op^W(e^{-\beta
q_m})}{\Tr Op^W
(e^{-{\beta}q_m})}=\int_{\Lambda_\infty} a\circ
\Pi_{m_1}\, d\mu_G{\beta}:=\tilde\omega_\beta (A)
\ee
and if moreover $\hat a$ and $\hat b$ are bounded
measures on $\Lambda_{m_1}$, (\ref{rmixing})
becomes:
\be
\label{rmixinga}
\lim_{t\to\infty}\lim_{m\to\infty}
\tilde\omega_\beta \big( A_m(t)B\big) =
\tilde\omega_\beta ( A)\cdot
\tilde\omega_\beta ( B)
\ee
\end{itemize}
\end{theorem}
\vskip 0.5cm\noindent
{\bf Remarks.}
\begin{enumerate}
\item Here, the choice of the quadratic form 
$\tilde q_{\beta ,m}$ is dictated from the fact
that we have (see Lemma \ref{lemma3} below):
\be
\label{2betaq} e^{-\tilde q_{\beta ,m}}\#
e^{-\tilde q_{\beta ,m}}= C_{\beta ,m}''e^{-2\beta
\tilde q_m}
\ee
where
$$
\tilde q_m(\xxi )= \frac12|\xi|^2+\la \tilde V_m
x,x\ra.
$$  
In view of assumption (H6)(i), this explains
why we get the usual Gibbs measure in the limit
$m\rightarrow +\infty$. Note that we also have
$\ds F_{\beta ,m}=\frac{\beta}2I_m+{\cal O}(\beta^3
)$ et $G_{\beta ,m}=\beta V_m+{\cal O}(\beta^3)$, so
that $\tilde q_{\beta ,m}$ is asymptotically equal
to
$\beta\tilde q_m$ as $\beta\rightarrow 0_+$. 
\item  For $\beta$ small it is possible to compare
$Op^W(e^{-\tilde q_{\beta ,m}})$ with
$e^{-\beta\tilde H_m}$, where $\ds\tilde
H_m=Op^W(\tilde
q_m)=\frac12(\sum_{j=1}^{2m+1}D_{x_j}^2)+\la \tilde
V_m x,x\ra$. Actually, denoting $\tilde\lambda_1,
\ldots, \lambda_{2m+1}$ the eigenvalues of $\tilde
V_m$, we get by standard formulas (see (\ref{L3})
below) that $Op^W(e^{-\tilde q_{\beta ,m}})$ is
unitarily equivalent to 
\be 
\label{R1}
\bigotimes_{j=1}^{2m+1}\frac1{\sqrt{1-\beta^2
\tilde\lambda_j/2}}{\rm
exp}\left[-\frac{1}{2}\left( {\rm
ln}\frac{1+\beta{\sqrt{\tilde\lambda_j/2}}}
{1-\beta{\sqrt{\tilde\lambda_j/2}}}\right)
(D_{x_j}^2+ x_j^2)\right] 
\ee
 while under the same unitary transformation
$e^{-\beta\tilde H_m}$ becomes
\be 
\label{R2}
\bigotimes_{j=1}^{2m+1}e^{-\beta{\sqrt
{\tilde\lambda_j/2}}(D_{x_j}^2+ x_j^2)}. 
\ee 
Since for small $\beta$ we have: 
$$
\frac{1}{2}{\rm
ln}\frac{1+\beta{\sqrt{\tilde\lambda_j/2}}}
{1-\beta{\sqrt{\tilde\lambda_j/2}}}
=\beta{\sqrt{\frac{\tilde\lambda_j}2}}\Big( 1+{\cal
O}(\beta)\Big) 
$$  
 we get from (\ref{R1})-(\ref{R2}):
\be
\label{R3} Op^W(e^{-\tilde q_{\beta ,m}})=
e^{-\beta\tilde H_m(1+\beta R_{\beta ,m})}
\ee
where $[R_{\beta ,m,}\tilde
H_m]=0$ and 
$\ds\frac1{2m+1}R_{\beta ,m}$ is uniformly
 bounded in $(\beta,m)$. 
\item The previous remark proves that for $\beta$
small, 
$[Op^W(e^{-\tilde q_{\beta ,m}}), \tilde H_m
]=0$. This is  true
for all positive 
$\beta$, by formula
(\ref{5}) below, and the fact that 
$$ e^{-\tilde q_{\beta ,m}}\circ {\rm
exp}tH_{\tilde q_m} (\xxi )= e^{-\tilde q_{\beta
,m}}\circ {\rm exp}tH_{\tilde  q_{\beta ,m}}\big(
(2F_{\beta ,m})^{-\frac12}x, (2F_{\beta
,m})^{-\frac12}\xi\big)
$$ is constant with respect to $t\in\R$.
\item  Since the choice of the lattice $\Z$ does
not play any role at all in our
 proofs, it can be replaced without modification by
any lattice 
$\Gamma\subset \R^d$ ($d\geq 1$) of the type
considered in \cite{LL}, in which case the model
describes an infinite harmonic crystal in $\R^d$.
Also the choice of 
$\Lambda_m$ is unessential, in the sense that any
other choice 
$\Lambda_m'\rightarrow \Lambda_\infty$ in a
reasonable way leads to the same results. 
\end{enumerate}

\vskip 1.0cm  
\section{Proof of Theorem 2.1} 
\setcounter{equation}{0}%
\setcounter{theorem}{0}%
\setcounter{proposition}{0}%
\setcounter{lemma}{0}%
\setcounter{corollary}{0}%
\setcounter{definition}{0}%

\par Now we turn to the proof of Theorem 2.1. The 
fact that $(\Lambda_\infty , \phi_t, \hat\mu_\beta
)$ is ergodic essentially follows from the arguments 
of \cite{LL}, but, for the sake of completeness,
we give a sketch of the proof. First, the invariance 
of $\hat\mu_{\beta}$ under $\phi_t$ is a
consequence of the commutativity between the operator $B$ defined in (\ref{A(xxi)}) and the
operator $\ds\ell^2(\Z )\oplus \ell^2(\Z )\ni 
(\xxi )\mapsto (W_\beta^{\frac12}x,
W_\beta^{\frac12}\xi )$. Now, denote
$\hat h_1$ the closure in
$L^2(\Lambda_\infty , d\hat\mu_\beta )$ of the set 
of all finite sums of $(a_jx_j+b_j\xi_j)$'s,
($a_j,b_j\in\C$ ;
$j\in\Z$). Then, denoting
$d(\Z )$ the elements of $\ell^2(\Z )$ with finite 
support, the application
\begin{eqnarray} 
\Theta\; :\; d(\Z ) \oplus d(\Z ) && 
\rightarrow \;\; \hat h_1\nonumber\\ 
a \oplus b && \mapsto \sum_j(a_jx_j+b_j\xi_j)
\nonumber
\end{eqnarray}
can be extended into an isomorphism from 
$D((2VW_\beta )^{-1/2}) \oplus D(W_\beta^{-1/2})$ to
$\hat h_1$ (where $D(A)$ denotes the domain of the
operator $A$). Moreover, identifying $D((2VW_\beta
)^{-1/2}) \oplus D(W_\beta^{-1/2})$ with $\ell^2(\Z )
\oplus \ell^2(\Z )$ in an obvious way, we see that
the action of $\phi_t$ on $\hat h_1$ is represented
on $\ell^2(\Z ) \oplus \ell^2(\Z )$  (via the two
previous identifications) by its infinitesimal
generator $\ds U=\left(\begin{array}{clcr} 0 &
-V^{-1/2}\\ V^{1/2} & \quad 0\end{array}\right)$.
Since, by assumption (H2), $U$ has no point spectrum
the result (i) follows by an abstract argument (see
\cite{LL}Prop.4.2).

\par To prove (ii) we first show :
\begin{lemma}
\label{lemma}
There exists a constant $C_{\beta ,m}$ such that :
$$
e^{-\beta H_m}=C_{\beta,m}Op^W(e^{-q_{\beta ,m}})
$$
\end{lemma}
{\it Proof} Let $\lambda_1,...,\lambda_{2m+1}$
 be the eigenvalues of $V_m$, and denote 
$y=(y_1,...,y_{2m+1})$ the coordinates in $\R^{2m+1}$ corresponding to an orthonormal basis of
eigenvectors of $V_m$. Then  $H_m$ becomes:
$$
H_m' = -\frac12\Delta_y+\sum_j\lambda_jy_j^2
$$
while the operator $K_m=Op^W(e^{-q_{\beta ,m}})$ 
is transformed into :
$$
K_m'=\bigotimes_{j=1}^{2m+1}Op^W\left(
\exp{-\Big(\frac1{\sqrt{2\lambda_j}}{\rm
tanh}\big(\beta\sqrt{\frac{\lambda_j}2}\big) 
\eta_j^2 +\sqrt{2\lambda_j}{\rm
tanh}\big(\beta\sqrt{\frac{\lambda_j}2}\big) 
y_j^2\Big)}\right).
$$ 
(Here $\eta$ is the dual variable of $y$, 
and   $\xi=(^tM)^{-1}\eta=M\eta$ since
 $x=My$ with $M$ orthogonal.)
Then the change of variables
$$
y_j\mapsto z_j=(2\lambda_j)^{1/4}y_j
$$ 
transforms
$H_m'$ into: 
\be
\label{L1}
H_m''=\sum_j\sqrt{\frac{\lambda_j}2}(D_{z_j}^2+z_j^2)
\ee
and $K_m'$ into :
\be
\label{L2}
K_m''=\bigotimes_{j=1}^{2m+1}Op^W\left( e^{ -{\rm
tanh}\big(\beta\sqrt{\frac{\lambda_j}2}\big)
\big( \zeta_j^2+z_j^2\big)}\right).
\ee
Next, consider the well known one-dimensional identity valid for $0<a<1$
\[ 
Op^W(e^{-a(x^2+\xi^2)})=\frac1{\sqrt{1-a^2}}
e^{-\frac{a}{2}x^2}
e^{-\frac{a}{1-a^2}D_x^2}e^{-\frac{a}{2}x^2}
\]
which can be for instance verified by 
explicit computation of the Weyl symbol of the r.h.s..
Then, using the formula (see e.g.\cite{He})
\[
 e^{-\frac{x^2}{2}} e^{-tD_x^2}e^{-\frac{x^2}{2}}= 
{\rm exp}\left[-\frac{{\rm
ln}(z+\sqrt{z^2-1})}{4k\sqrt{z^2-1}}
(D_x^2+4k^2(z^2-1)x^2)\right]
\]
with ${\ds k=\frac1{4t}}$, $z=2t+1$ $(t>0)$, 
we get in particular for $0<a<1$ :
\be\label{L3} Op^W(e^{-a(x^2+\xi^2)})=
\frac1{\sqrt{1-a^2}}{\rm exp}\left[-\frac{1}{2}
\left(
{\rm ln}\frac{1+a}{1-a}\right)(D_x^2+x^2)\right]
\ee
Taking $a={\rm tanh}\big(\beta\sqrt{\frac{\lambda_j}2}
\big)$, the Lemma follows from
(\ref{L1})-(\ref{L3}).
\hfill$\square$
\vskip 0.4cm
 Now, since the flow generated by $q_n$ 
defines a linear
canonical transformation on $\Lambda_n$, we have:
\be
\label{5} 
e^{itH_n}Op^W(a)e^{-itH_n}= Op^W(a\circ \exp
tH_{q_n}), \qquad \forall a\in S_n(1)
\ee
This relation (an ``exact Egorov theorem",
 going back at least to Van Hove (see
e.g.\cite{Fo}))  holds only in the Weyl
quantization
\cite{BS}\S 5.2. 

\par For $n\leq m$ and $(\xxi )\in \Lambda_m$, denote
\begin{eqnarray} 
\label{6}
&&\rho (m, \lambda )=\| e^{-\frac12\beta H_m}f_\lambda\|^2\\
&&a_{n,T}(\xxi )=\frac1T\int_0^T 
(a\circ\Pi_{m_1}\circ{\rm
exp}tH_{q_n}\circ\Pi_n)(\xxi ))dt. 
\end{eqnarray}
Using Lemma \ref{lemma}, (\ref{5}) and (\ref{w_f}), 
we get:
\be
\label{7}
A(m,n,T,\lambda )=\frac{C_{\beta ,m}^2}
{\rho (m,\lambda )}\int_{\Lambda_m}\left( e^{-
q_{\beta ,m}}\# a_{n,T}\# e^{- q_{\beta ,m}}\right) 
(\xxi )w_{f_\lambda}(\xxi )dxd\xi
\ee
where $\#$ is the Weyl composition of symbols 
on $\Lambda_m$:
\be
\label{8}
(a\# b)(\xxi )=\pi^{-2(2m+1)}
\int_{\Lambda_m^2} a(x+y,\xi +\eta )b(x+z ,\xi +\zeta
) e^{2i(\zeta y-z\eta )}dyd\eta dzd\zeta .
\ee
Taking advantage of assumption (H3), we get
from (\ref{7}) 
\begin{eqnarray}
\label{9}
\int_{X_m}A(m,n,T,\lambda )\rho (m,\lambda )
d\theta_m (\lambda )=C_0\int_{\Lambda_m}
\left( e^{-q_{\beta ,m}}\# a_{n,T}\# e^{-q_{\beta ,m}}
\right) (\xxi )dxd\xi\; \\
\label{9'}\int_{X_m}\vert A(m,n,T,\lambda )
\vert \rho (m,\lambda )d\theta_m (\lambda )\leq
C_0\int_{\Lambda_m}
\vert\left( e^{-q_{\beta ,m}}\# a_{n,T}
\# e^{-q_{\beta ,m}}\right) (\xxi )\vert dxd\xi\; 
\end{eqnarray}
where $C_0=C_0(\beta ,m)$ is a constant, which 
can be computed by taking $a\equiv 1$ in (\ref{9}):
\be
\label{10}
C_0 = \left( \int  e^{- q_{\beta ,m}}\# 
e^{-q_{\beta ,m}}dxd\xi\right)^{-1}\int_{X_m}\rho
(m,\lambda )d\theta_m (\lambda ).
\ee
This also proves (\ref{fini}) so that, using 
the notation (\ref{nutilde}) we can rewrite
(\ref{9})-(\ref{9'}) as:
\begin{eqnarray}
\label{11}
\int_{X_m}A(m,n,T,\lambda )d\nu_m(\lambda )=
C_1\int_{\Lambda_m}
\left( e^{-q_{\beta ,m}}\# a_{n,T}\#
 e^{-q_{\beta ,m}}\right) (\xxi )dxd\xi\; \\
\label{11'}\int_{X_m}\vert A(m,n,T,\lambda )
\vert d\nu_m(\lambda )\leq
C_1\int_{\Lambda_m}
\vert\left( e^{-q_{\beta ,m}}\# a_{n,T}\# 
e^{-q_{\beta ,m}}\right) (\xxi )\vert dxd\xi\; 
\end{eqnarray}
with 
$$
C_1=\left( \int  e^{- q_{\beta ,m}}\# e^{-q_{\beta
,m}}dxd\xi\right)^{-1}.
$$ 
Now we make use of the two
following properties of the operation $\#$ (valid
e.g. for any $a$, $b$, $c$ in ${\cal
S}(\Lambda_m)$):
 \begin{eqnarray} 
\label{12} 
&&
\int_{\Lambda_m} (a\# b\# c)(\xxi )dxd\xi = 
\int_{\Lambda_m} (b\# c\# a)(\xxi )dxd\xi\\
\label{12'}
&&
\int_{\Lambda_m} (a\# b)(\xxi )dxd\xi = 
\int_{\Lambda_m} a(\xxi )b(\xxi )dxd\xi.
\end{eqnarray}
The property (\ref{12}) is just a consequence of 
the cyclicity of the trace of operators, and
(\ref{12'}) comes from a direct computation 
using (\ref{8}). 
\par Here our symbol $a$ is not supposed to be in 
${\cal S}(\Lambda_m)$, but an easy argument of
density allows us to deduce from (\ref{11}) and 
(\ref{12})-(\ref{12'}) (using also (\ref{2beta})):
\be
\label{13}
\int_{X_m}A(m,n,T,\lambda )d\nu_m(\lambda )=
\int_{\Lambda_m}
a_{n,T}(\xxi )\left[ e^{-q_{\beta ,m}(\xxi )}
dxd\xi\right]_N
\ee
where we have used the notation:
\be
\label{norm}
\int_E f\left[ d\mu\right]_N = 
\frac1{\mu(E)}\int_E fd\mu
\ee
for any finite positive measure $\mu$ on a set $E$. 
\par Now the problem is to rewrite also (\ref{11'}) 
in this way, despite the appearance of the
modulus. The argument to do this is based upon
the following: 
\begin{lemma}
\label{lemma2}
There exists a positive definite quadratic form 
$Q_{\beta ,m}(\xxi ,y,\eta )$ on $\Lambda_m^2$ such
that for all $a\in S_m(1)$:
$$
e^{-q_{\beta ,m}}\# a\# e^{-q_{\beta ,m}}=C_{\beta
,m}'\tilde a e^{-q_{2\beta ,m}}$$ where $C_{\beta
,m}'$ is the constant appearing in (\ref{2beta}), and
$$
\tilde a(\xxi )=\int_{\Lambda_m}a(y,\eta )\left[
e^{-Q_{\beta ,m}(\xxi ,y,\eta )}dyd\eta\right]_N.
$$
\end{lemma} 
{\it Proof} - See Appendix 1.
\vskip 0.4cm
We deduce in particular from Lemma \ref{lemma2} 
the existence of a positive $C^\infty$ function
$\gamma (x,\xi )$ on $\Lambda_m$ such that for all 
$a\in S_m(1)$:
\begin{eqnarray}
\label{14} 
&&\int_{\Lambda_m}\big( e^{-q_{\beta ,m}}\# a\# 
e^{-q_{\beta ,m}}\big) dxd\xi =
\int_{\Lambda_m}a(x,\xi )\gamma (x,\xi )dxd\xi \\
\label{14'} 
&&\int_{\Lambda_m}\vert e^{-q_{\beta ,m}}\# a\# 
e^{-q_{\beta ,m}}\vert dxd\xi \leq
\int_{\Lambda_m}\vert a(x,\xi )\vert\gamma 
(x,\xi )dxd\xi .
\end{eqnarray}
\par By (\ref{11}), (\ref{13}) and (\ref{14}) 
we get that $\gamma$ equals a constant times
$e^{-q_{2\beta ,m}}$, which by (\ref{11'}) and 
(\ref{14'}) allows us to conclude that :
\be
\label{15}
\int_{X_m}\left\vert A(m,n,T,\lambda )
\right\vert d\nu_m(\lambda )\leq\int_{\Lambda_m}
\vert a_{n,T}(\xxi )\vert\left[ 
e^{-q_{\beta ,m}(\xxi )}dxd\xi\right]_N.
\ee
\par Without loss of generality, we can assume 
from now on that $\ds\int_{\Lambda_\infty} (a\circ
\Pi_{m_1})d\hat\mu_{\beta}=0$, and then it 
remains to estimate the r.h.s. of (\ref{15}). Since
$a_{n,T}(\xxi )$ depends only on $\Pi_n(\xxi )$, 
we can let $m$ go to $+\infty$ in (\ref{15}) and
we get:
\be
\label{16}
\limsup_{m\to\infty}\int_{X_m}\left\vert
 A(m,n,T,\lambda )\right\vert d\nu_m(\lambda )
\leq\int_{\Lambda_\infty}
\vert a_{n,T}(\xxi )\vert d\hat\mu_{\beta}.
\ee
\par Then we use  (\ref{1}) to let $n$ go to 
$+\infty$ in
(\ref{17}). By the dominated convergence theorem, 
we then obtain:
\be
\label{17}
\limsup_{n\to\infty}\limsup_{m\to\infty}
\int_{X_m}\left\vert A(m,n,T,\lambda )\right\vert
d\nu_m(\lambda )
\leq\int_{\Lambda_\infty}
\vert \frac1T\int_0^T(a\circ\Pi_{m_1}
\circ\phi_t)(\xxi )dt\vert d\hat\mu_{\beta}.
\ee
Finally, we let $T$ go to $+\infty$. 
By the ergodicity property, we have that
$$
\frac1T\int_0^T(a\circ\Pi_{m_1}\circ\phi_t)(\xxi )dt
\rightarrow \int_{\Lambda_\infty} (a\circ
\Pi_{m_1})d\hat\mu_{\beta}=0
$$ 
for $\hat\mu_{\beta}$-almost all $(\xxi )$ 
in $\Lambda_\infty$.
Therefore, applying again the dominated 
convergence theorem, we get from (\ref{17}):
\be
\label{18}
\lim_{T\to +\infty}\limsup_{n\to\infty}
\limsup_{m\to\infty}\int_{X_m}\left\vert
A(m,n,T,\lambda )\right\vert d\nu_m(\lambda )
\leq 0
\ee
and this completes the proof of Theorem 2.1.
\hfill$\square$
\vskip 1.0cm  
\section{Proof of Theorem 2.2} 
\setcounter{equation}{0}%
\setcounter{theorem}{0}%
\setcounter{proposition}{0}%
\setcounter{lemma}{0}%
\setcounter{corollary}{0}%
\setcounter{definition}{0}%
Let us now proceed to the proof of Theorem 2.2.
Denote
\be
\omega_{\beta ,m}(A)=\frac{\Tr (Ae^{-\beta
H_m})}{\Tr (e^{-\beta H_m})}.
\ee 
Using Lemma \ref{lemma} and (\ref{12'}) we see
that
\be
\omega_{\beta ,m}(A)=\int_{\Lambda_m} a\circ
\Pi_{m_1}(\xxi)
\left[e^{-q_{\beta ,m} (\xxi)}\,dx\,d\xi\right]_N
\ee 
so that the first assertion (\ref{th3.2a}) of
the theorem is obvious.

For $m\geq n\geq m_1$ we also have:
\be
\label{omegam}
\omega_{\beta ,m}(A_n(t)B) =\int_{\Lambda_m}
a_{n,t}\# (b\circ \Pi_{m_1})(\xxi)
\left[e^{-q_{\beta ,m} (\xxi)}\,dx\,d\xi\right]_N.
\ee 
For $X=(x,\xi )$ and $Y=(y,\eta )\in\Lambda_m$,
we denote 
\be
\sigma (X,Y) = \xi y -x\eta
\ee the canonical symplectic form on $\Lambda_m$.
Then by (\ref{8}) we have:
\be
\label{diese} a_{n,t}\# (b\circ \Pi_{m_1})(X)
=\pi^{-2(2m+1)}\int_{\Lambda_n^2} (a_{n,t}\circ
\Pi_n)(Y)(b\circ \Pi_{m_1})(Z) e^{2i[\sigma
(Y,X)+\sigma (Z,Y-X)]} dYdZ
\ee 
By the Fourier inversion formula and the
assumption on
$a, b$, we can write for any
$Y_1\in\Lambda_{m_1}$:
\be
\label{fourierinv} a(Y_1)=(2\pi )^{-2(2m_1+1)}
\int_{\Lambda_{m_1}} e^{\la Y_1,Y^*\ra}\hat
a(Y^*)dY^*
\ee and a similar formula for $b$. Here we have
used an abuse of notation by writing
$\ah(Y^*)dY^*$ for the (non necessarily Lebesgue
absolutely continuous) measure defined by the
Fourier transform of $a$.

In particular, taking $Y_1=\Pi_{m_1}\phi_{n,t}
\Pi_n(Y)$ in (\ref{fourierinv}) and substituing in
(\ref{diese}), we get:
\begin{eqnarray*} && a_{n,t}\# (b\circ \Pi_{m_1})(X)
=\pi^{-2(2m+1)}(2\pi )^{-4(2m_1+1)}\times
\\ &&\qquad\qquad \int e^{2i[\sigma (Y,X)+\sigma
(Z,Y-X)]+i\la
\phi_{n,t}\Pi_n(Y),Y^*\ra +i\la Z,Z^*\ra}
 \hat a(Y^*)dY^*\, \hat b(Z^*)dZ^*\,dYdZ
\end{eqnarray*} where the integration runs over
$(Y^*,Z^*,Y,Z)\in \Lambda_{m_1}\times
\Lambda_{m_1}\times\Lambda_m\times\Lambda_m$, and
$\Lambda_{m_1}$ has been identified in an obvious
way with a subspace of $\Lambda_{n}$ and of
$\Lambda_{m}$.

Interpreting the integration over $(Y,Z)$ as an
oscillatory one, we can first integrate with
respect to
$Z$, and we obtain (using the well-known identity
$\int_{\R^d}e^{2i(x-y)\xi}d\xi = \pi^d
\delta (y=x)$):
\be
\label{diesebis} a_{n,t}\# (b\circ \Pi_{m_1})(X)
=(2\pi )^{-4(2m_1+1)}\int e^{i\la Z^*,X\ra +i\la
\phi_{n,t}\Pi_n(X+\tilde Z^*/2),Y^*\ra}
 \hat a(Y^*)dY^*\, \hat b(Z^*)dZ^*
\ee where we have denoted $\tilde Z^*= (-\zeta^*,
z^*)$ if $Z^*=(z^*,\zeta^*)$.

Now, inserting (\ref{diesebis}) into
(\ref{omegam}), and making the change of variables
$X\mapsto X-\tilde Z^*/2$, this gives:
\begin{eqnarray*} 
&&\omega_{\beta ,m}(A_n(t)B)= (2\pi
)^{-4(2m_1+1)}\times 
\\ && \quad 
\int_{\Lambda_m\times\Lambda_{m_1}^2} e^{i\la
Z^*,X\ra +i\la
\phi_{n,t}\Pi_n(X),Y^*\ra}
\left[e^{-q_{\beta ,m} (X-\tilde Z^*/2)}dX\right]_N
 \hat a(Y^*)dY^*\, \hat b(Z^*)dZ^*.
\end{eqnarray*} and therefore, writing $q_{\beta
,m}(X)=\la Q_{\beta ,m}X,X\ra$ with
$Q_{\beta ,m}(x,\xi )=(V_mW_{\beta ,m}x\, ,\,
W_{\beta ,m}\xi )$:
\be
\label{"}
\omega_{\beta ,m}(A_n(t)B)=(2\pi )^{-4(2m_1+1)}
\int_{\Lambda_{m_1}^2}
\Gamma_{m,n,t} (Y^*,Z^*)e^{-q_{\beta ,m}(\tilde Z^*
)/4}
 \hat a(Y^*)dY^*\, \hat b(Z^*)dZ^*.
\ee where
\be
\label{Gamma_m}
\Gamma_{m,n,t} (Y^*,Z^*) =\int_{\Lambda_m} e^{i\la
Z^*,X\ra +\la\tilde Z^*,Q_{\beta ,m}X\ra +i\la
\phi_{n,t}\Pi_n(X),Y^*\ra}
\left[e^{-q_{\beta ,m} (X)}dX\right]_N
\ee is of the form:
\be
\label{F_m}
\Gamma_{m,n,t} (Y^*,Z^*) =\int_{\Lambda_m}
F_{n,t,Y^*,\tilde Z^*}(\Pi_{m_1}Q_{\beta ,m}X,
\Pi_{n}X))\,
\left[e^{-q_{\beta ,m} (X)}dX\right]_N
\ee with $F_{n,t,Y^*,Z^*}$ smooth and uniformly
bounded together with all its derivatives on
$\Lambda_{m_1}\times
\Lambda_n$.
To let $m$ tend to infinity in (\ref{F_m}), we use
the following lemma (which is the point where (H5)
is used):

\begin{lemma}
\label{limit} Let $F\in
C^{\infty}(\Lambda_{m_1}\times
\Lambda_n)$ be uniformly bounded together with all
its derivatives. Then
$$
\int_{\Lambda_m} F(\Pi_{m_1}Q_{\beta ,m}X,
\Pi_{n}X))\,
\left[e^{-q_{\beta ,m} (X)}dX\right]_N\rightarrow
\int_{\Lambda_\infty} F(\Pi_{m_1}Q_{\beta}X,
\Pi_{n}X))\, d\hat\mu_\beta\quad
(m\rightarrow\infty )
$$ 
where $\hat\mu_\beta$ is
defined in {\rm (\ref{muhat})}, and $Q_\beta$ is
defined on
$\Lambda_\infty$ by:
$$
Q_\beta (x,\xi )=(VW_\beta x,W_\beta\xi ), \qquad 
W_\beta \sqrt2 V^{-\frac12}{\rm tanh}\frac{\beta
V^{\frac12}}{\sqrt2}.
$$
\end{lemma} {\it Proof}- See Appendix 1.
\vskip 0.5cm\noindent 
Now, for any fixed $(n,t,Y^*,Z^*)$, we
see on (\ref{Gamma_m})-(\ref{F_m}) that, as
$m\rightarrow
\infty$,
$\Gamma_{m,n,t}(Y^*,Z^*)$ tends to:
\be
\label{Gamma_n}
\Gamma_{n,t}(Y^*,Z^*) =\int_{\Lambda_\infty}
e^{i\la Z^*,X\ra +\la\tilde Z^*,Q_{\beta }X\ra +i\la
\phi_{n,t}\Pi_n(X),Y^*\ra}d\hat\mu_\beta
\ee which in turns is of the form:
\be
\label{fg}
\Gamma_{n,t}(Y^*,Z^*) =\int_{\Lambda_\infty}
f_{Z^*}(X)g_{Y^*}(\Pi_{m_1}\phi_{n,t}\Pi_nX)
d\hat\mu_\beta
\ee with $f_{Z^*}$ and $g_{Y^*}$ uniformly bounded,
and $g_{Y^*}$ continuous on $\Lambda_{m_1}$. Then,
using (\ref{1}) and the dominated convergence
theorem, we see on (\ref{fg}) that, as
$n\rightarrow\infty$, $\Gamma_{n,t}(Y^*,Z^*)$ tends
to:
\be
\label{Gamma_t}
\Gamma_t(Y^*,Z^*)=
\int_{\Lambda_\infty}
f_{Z^*}(X)g_{Y^*}(\Pi_{m_1}\phi_{t}X)
d\hat\mu_\beta .
\ee Now, the same arguments used in the proof of
Theorem 2.1 (i) (see the begining of section 3)
lead to the fact that under (H4) the classical
dynamical system $(\Lambda_\infty,
\phi_t, \hat\mu_\beta )$ is mixing. As a
consequence, we get from (\ref{Gamma_t}):
\be
\label{Gamma}
\Gamma_t(Y^*,Z^*)\rightarrow
\int f_{Z^*}(X)d\hat\mu_\beta
\cdot
\int g_{Y^*}(\Pi_{m_1}X)d\hat\mu_\beta
\quad {\rm as}\quad t\rightarrow\infty .
\ee Summing up (\ref{Gamma_m})-(\ref{Gamma}), we
have proved that for any fixed
$(Y^*,Z^*)\in\Lambda_{m_1}^2$, we have:
\be
\label{Gamma'}
\lim_{t\rightarrow\infty}
\lim_{n\rightarrow\infty}
\lim_{m\rightarrow\infty}
\Gamma_{m,n,t}(Y^*,Z^*) =
\int_{\Lambda_\infty} e^{i\la Z^*,X\ra +\la\tilde
Z^*,Q_{\beta }X\ra}d\hat\mu_\beta (X)
\cdot
\int_{\Lambda_\infty} e^{i\la
X,Y^*\ra}d\hat\mu_\beta (X)
\ee and because of the translation invariance of
the Lebesgue measure on $\Lambda_m$, and the fact
that $\la Z^*,\tilde Z^*\ra=0$, it is also easy to
verify that
\be
\label{ch}
\int_{\Lambda_\infty} e^{i\la Z^*,X\ra +\la\tilde
Z^*,Q_{\beta }X\ra}d\hat\mu_\beta (X) =e^{\la
Q_\beta \tilde Z^*,\tilde Z^*\ra /4}
\int_{\Lambda_\infty} e^{i\la
Z^*,X\ra}d\hat\mu_\beta (X)
\ee

 Since by assumption
$\hat a(Y^*)dY^*$ and
$\hat b(Z^*)dZ^*$ are bounded measures on
$\Lambda_{m_1}$, and $\vert
\Gamma_{m,n,t} (Y^*,Z^*)e^{-q_{\beta ,m}(\tilde Z^*
)/4}\vert =1$, we can use the dominated convergence
theorem in (\ref{"}) and conclude from
(\ref{Gamma'})-(\ref{ch}) (using also the obvious
fact that $q_{\beta ,m}(\tilde Z^*)$ tends to $\la
Q_\beta \tilde Z^*,\tilde Z^*
\ra$ as $m\rightarrow\infty$) that:
\begin{eqnarray*} &&\lim_{t\rightarrow\infty}
\lim_{n\rightarrow\infty}
\lim_{m\rightarrow\infty}
\omega_{\beta ,m}(A_n(t)B) =(2\pi
)^{-4(2m_1+1)}\times \\ &&\quad
\int_{\Lambda_{m_1}\times\Lambda_\infty} e^{i\la
X,Y^*\ra} \hat a(Y^*)dY^*\, d\hat\mu_\beta (X)
\cdot
\int_{\Lambda_{m_1}\times\Lambda_\infty} e^{i\la
Z^*,X\ra}
\hat b(Z^*)dZ^*\, d\hat\mu_\beta (X) \\ &&\quad
=\int_{\Lambda_\infty}a\circ \Pi_{m_1}
d\hat\mu_\beta \cdot
\int_{\Lambda_\infty}b\circ \Pi_{m_1} d\hat\mu_\beta
\end{eqnarray*} where the last equality comes again
from the Fourier-inverse formula. \hfill$\square$.
\vskip 1.0cm\noindent
\section{Proof of Theorem 2.3}
\begin{lemma}
\label{lemma3}
For any pair of positive definite real-symmetric 
matrices $F$ and $G$ on $\Lambda_m$, there exists a
constant
$C=C(F,G,m)$ such that:
$$
e^{-(\la F\xi ,\xi\ra +\la Gx,x\ra)}\# 
e^{-(\la F\xi ,\xi\ra +\la Gx,x\ra)}
=Ce^{-2\big( \la F(F+G^{-1})^{-1}G^{-1}
\xi ,\xi\ra +
 \la (F+G^{-1})^{-1}x ,x\ra\big)}.
$$
\end{lemma}
{\it Proof} - See Appendix 1.
\vskip 0.4cm
In particular, taking $F=F_{\beta ,m}$ and  
$G=G_{\beta ,m}$ defined in
(\ref{F})-(\ref{G}) we get easily (\ref{2betaq}) 
from Lemma \ref{lemma3}. Then 
computations analogous to those of the previous 
section ((\ref{7}) through (\ref{15})) lead to:
\be
\label{19}
\int_{X_m}\left\vert \tilde A(m,T,\lambda )
\right\vert d\tilde\nu_m(\lambda
)\leq\int_{\Lambda_m} \vert a_{m,T}(\xxi )\vert\left[
e^{-\beta \tilde q_{m}(\xxi )}dxd\xi\right]_N. 
\ee
Now in the r.h.s. (\ref{19}) we make the change 
of variables :
$$
x = \tilde V_m^{-\frac12}y \qquad \xi =\eta 
$$
which gives :
\be
\label{20}
\int_{\Lambda_m}
\vert a_{m,T}(\xxi )\vert\left[ e^{-\beta \tilde
q_{m}(\xxi )}dxd\xi\right]_N = \int_{\Lambda_m}
\vert a_{m,T}(\tilde V_m^{-\frac12}y,\eta )
\vert\left[ e^{-\beta (\eta^2 +y^2)}dyd\eta\right]_N.
\ee
Since the quadratic form in the exponent is 
now diagonal, we can integrate over $n\geq m$
variables so that
\be
\label{21}
\int_{\Lambda_m}
\vert a_{m,T}(\xxi )\vert\left[ e^{-\beta 
\tilde q_{m}(\xxi )}dxd\xi\right]_N
= \int_{\Lambda_n}
\vert a_{m,T}(\tilde V_m^{-\frac12}\Pi_m y,
\Pi_m\eta )\vert\left[ e^{-\beta (\eta^2
+y^2)}dyd\eta\right]_N
\ee
for any $n\geq m$. Coming back to the 
old variables on $\Lambda_n$, this gives:
\be
\label{22}
\int_{\Lambda_m}
\vert a_{m,T}(\xxi )\vert\left[ e^{-\beta \tilde
q_{m}(\xxi )}dxd\xi\right]_N = \int_{\Lambda_n}
\vert a_{m,T}(\tilde V_m^{-\frac12}\Pi_m 
\tilde V_n^{\frac12}x,\Pi_m\xi )\vert\left[ 
e^{-\beta\tilde q_{n}(\xxi
)}dxd\xi\right]_N.
\ee
Now, by assumption (H4)(ii), for $n$ large enough, 
the function
$\ds a_{m,T}(\tilde V_m^{-\frac12}\Pi_m 
V_n^{\frac12}x,\Pi_m\xi )$ depends only on a fixed
number of variables independent of $n$. 
Then, by assumption (H4)(i) and standard results on
the Gaussian measures (see e.g. \cite{LL}), 
letting $n$ tend to $+\infty$ in
(\ref{22}), we get:
\be
\label{23}
\int_{\Lambda_m}
\vert a_{m,T}(\xxi )\vert\left[ 
e^{-\beta\tilde q_{m}(\xxi )}dxd\xi\right]_N
= \int_{\Lambda_\infty}
\vert a_{m,T}(\tilde V_m^{-\frac12}
\Pi_m \tilde V_n^{\frac12}\Pi_n x,\Pi_m\xi )\vert
d\mu_G(\beta ). 
\ee
Finally, using assumption (H4)(iii), (\ref{1}), 
and the uniform (with respect to $m\geq
0$) continuity of exp$tH_{q_m}\circ
\Pi_m$ on each ${\cal H}_k$, we see that for
all $(\xxi )\in\Lambda_\infty$ and $t\in\R$:
\be
\label{24}
(a\circ \Pi_{m_1}\circ {\rm exp}tH_{q_m})
(\tilde V_m^{-\frac12}\Pi_m \tilde
V_n^{\frac12}\Pi_n x,\Pi_m\xi )
\rightarrow (a\circ \Pi_{m_1}\circ\phi_t)(\xxi )\quad
{\rm as}\quad m\rightarrow +\infty .
\ee
\par It follows from (\ref{24}), (\ref{23}), 
(\ref{19}) and the dominated convergence theorem
that:
\be
\label{25}
\limsup_{m\rightarrow +\infty}\int_{X_m}
\left\vert \tilde A(m,T,\lambda )\right\vert
d\tilde\nu_m(\lambda )\leq\int_{\Lambda_\infty}
\left\vert \frac1T\int_0^T(a\circ \Pi_{m_1}
\circ\phi_t)(\xxi )dt\right\vert d\mu_G(\beta ).
\ee
Letting $T$ tend to infinity in (\ref{25}), 
the ergodicity of the system 
$(\Lambda_\infty ,\phi_t,
\mu_G(\beta ))$ and the fact that we can restrict to
the case $\ds\int a\circ\Pi_{m_1}d\mu_G(2\beta )=0$
yield the assertion. \par$\hfill\hfill\square$
 \vskip 0.5cm   
\section{Coherent States: Sharpening the Ergodicity 
Result} 
\setcounter{equation}{0}%
\setcounter{theorem}{0}%
\setcounter{proposition}{0}%
\setcounter{lemma}{0}%
\setcounter{corollary}{0}%
\setcounter{definition}{0}%
 
\par Now we take $X_m=\Lambda_m$, 
$d\theta_m (\lambda 
)=d\lambda$, and for $\lambda =(\lambda_x
,\lambda_\xi )\in\Lambda_m$ the coherent states 
 defined by:
\be
\label{CS}
 f_\lambda(\xxi )=e^{ix\lambda_\xi - 
(x-\lambda_x)^2/2}.
\ee 
Then a direct computation gives:
\be 
\label{CS1}
w_{f_\lambda}(\xxi )=2^{2m+1}\pi^{m+\frac12
}e^{-(\xi -\lambda_\xi )^2-(x-\lambda_x)^2}
\ee so that (H3) is obviously satisfied. Therefore 
the results of Theorems 2.1 and 2.2 hold for
$V$ satisfying (H1)-(H2) (respectively (H1),
(H2) and (H6)).  The new fact which appears in this
situation is :
\begin{lemma}
\label{lemma0} Under (\ref{CS}), the two 
measures $d\nu_m(\lambda )$ and
$d\tilde\nu_m(\lambda )$, defined respectively by
(\ref{nutilde}) and (\ref{nutilde'}), are Gaussian
probability measures on $X_m=\Lambda_m$. \end{lemma}
{\it Proof} - In each case, the measure is of the
form $C\| Op^W(e^{-q})f_\lambda\|^2d\lambda$ where
$C$ is a constant and $q$ is a positive definite
quadratic form on $\Lambda_m$. Moreover, by
computations analogous e.g. to those for (\ref{7}) we
have : 
\be 
\|Op^W(e^{-q})f_\lambda\|^2=C'\int_{\Lambda_m}
(e^{-q}\#
e^{-q})(\xxi )w_{f_\lambda}(\xxi )dxd\xi 
\ee
 where
$C'$ is another constant. Then the result follows
immediately by (\ref{CS1}) and the fact that
$e^{-q}\# e^{-q}=C''e^{-q'}$ where $q'$ is a positive
definite quadratic form on $\Lambda_m$ and $C''$ is a
constant. \hfill$\square$ \vskip 0.4cm Now denote
$L_m$ and $\tilde L_m$ the two real-symmetric
positive definite $(4m+2)\times (4m+2)$-matrices
defined by : \begin{eqnarray} && d\nu_m(\lambda
)=\left[ e^{-\la L_m\lambda ,\lambda\ra}d\lambda
\right]_N\\ && d\tilde\nu_m(\lambda )=\left[ e^{-\la
\tilde L_m\lambda ,\lambda\ra}d\lambda \right]_N.
\end{eqnarray} Then for any bounded function
$A(\lambda )$ we have : \be   \int_{\Lambda_m}
A(\lambda )d\nu_m(\lambda
)=\int_{\Lambda_m}A(L_m^{-\frac12}\lambda )\left[
e^{-\vert\lambda\vert^2}d\lambda \right]_N  \ee and
therefore, by an argument similar to the one leading
to (\ref{23}) : \be \label{nuG}   \int_{\Lambda_m}
A(\lambda )d\nu_m(\lambda
)=\int_{\Lambda_\infty}A(L_m^{-\frac12}\Pi_m\lambda
)d\nu_G(\lambda )  \ee where $d\nu_G(\lambda )$ is
the infinite dimensional Gibbs measure obtained by
taking the limit of $\left[
e^{-\vert\lambda\vert^2}d\lambda \right]_N $ on
$\Lambda_n$ as $n\rightarrow +\infty$. A formula
analogous to (\ref{nuG}) is also true for
$d\tilde\nu_m(\lambda )$, and therefore it follows
from Theorems 2.1 and 2.2 that we have in this
situation : 
\begin{eqnarray} 
\label{CS2}
\lim_{T\to\infty}\limsup_{n\to\infty}
\limsup_{m\to\infty}\int_{\Lambda_\infty}\left\vert
A(m,n,T,L_m^{-\frac12}\Pi_m\lambda
)-\int_{\Lambda_\infty} a\circ \Pi_{m_1}\,
d\hat\mu_{\beta}\right\vert d\nu_G(\lambda
)=0\hfill\\  \label{CS3}
\lim_{T\to\infty}\limsup_{m\to\infty}
\int_{\Lambda_\infty}\left\vert
\tilde A(m,T,\tilde L_m^{-\frac12}\Pi_m\lambda
)-\int_{\Lambda_\infty} a\circ \Pi_{m_1}\,
d\mu_G({\beta})\right\vert d\nu_G(\lambda )=0.\hfill
\end{eqnarray} 
\par Finally, using a very
standard argument of measure theory, we easily deduce
from (\ref{CS2}) and (\ref{CS3}) the following:
\begin{proposition} 
Assume (H1)-(H2) and choose the
set of coherent states (\ref{CS}). Then there exist
sequences $(T_k)_{k\in\N}$, $(m_k)_{k\in\N}$,
$(n_k)_{k\in\N}$ simultaneously tending to $+\infty$
such that for $\nu_G$-almost all
$\lambda\in\Lambda_\infty$:
$$
\lim_{k\rightarrow+\infty}A(m_k,n_k,T_k,L_{m_k}^{-
\frac12}\Pi_{m_k}\lambda
) =\int_{\Lambda_\infty} a\circ \Pi_{m_1}\,
d\hat\mu_{\beta}.
$$ 
\par 
If moreover (H3) is
satisfied, there exist sequences $(T_k)_{k\in\N}$,
$(m_k)_{k\in\N}$ both tending to $+\infty$ such that
for $\nu_G$-almost all $\lambda\in\Lambda_\infty$ :
$$
\lim_{k\rightarrow+\infty}\tilde A(m_k,T_k,\tilde
L_{m_k}^{-\frac12}\Pi_{m_k}\lambda )
=\int_{\Lambda_\infty} a\circ \Pi_{m_1}\,
d\mu_G({\beta}).
$$ 
\end{proposition} 
{\bf Remarks.}
\begin{enumerate} \item 
An analogous result  holds
if (\ref{CS}) is replaced by the more
general case $\ds f_\lambda(\xxi
)=e^{ix\lambda_\xi - \la
F_m(x-\lambda_x),x-\lambda_x\ra}$, 
$F_m$ being any positive definite symmetric
matrix.  \item  Actually, one can replace
$L_m$ by any other symmetric matrix $L'_m$ such that
 $K_m=(L_m')^{-1/2}L_m(L_m')^{-1/2}$ is
a diagonal matrix and the measure $\ds
\left[ e^{-\la K_m\lambda
,\lambda\ra}d\lambda\right]_N$ on
$\Lambda_m$ admits a limit  $d\nu_\infty$
as $m\rightarrow +\infty$. In this case the
``$\nu_G$-almost all $\lambda$'' of the
Proposition must be replaced by
``$\nu_\infty$-almost all $\lambda$''. 
\end{enumerate} 

 \vskip 1.0cm      
\section{Appendix 1} 
\setcounter{equation}{0}%
\setcounter{theorem}{0}%
\setcounter{proposition}{0}%
\setcounter{lemma}{0}%
\setcounter{corollary}{0}%
\setcounter{definition}{0}%

{\bf 1. Proof of Lemma \ref{lemma2}}

\par Using (\ref{8}), we see that $e^{-q_{\beta ,m}}
\# a\# e^{-q_{\beta ,m}}$ can be put under the form:
\be
\label{A1} (e^{-q_{\beta ,m}}\# a\# e^{-q_{\beta
,m}}) (\xxi )=C_1\int_{\Lambda_m^4} a(Y_1
)e^{-q_1(x,\xi ,Y_1,Y_2, Y_3, Y_4)}dY_1dY_2dY_3dY_4
\ee
where all along this proof $C_j$ ($j=1,2...$) 
will denote  complex constants, $q_1$ is a  complex
quadratic form on $\Lambda_m^5$,  and the integral
(\ref{A1}) is oscillatory. Moreover, a direct
computation gives:
\be
\label{A2}
\int_{\Lambda_m^3}e^{-q_1(x,\xi ,Y_1,Y_2, Y_3,
Y_4)}dY_2dY_3dY_4 =  C_2e^{-Q(x,\xi ,Y_1 )}
\ee
where $C_2\in\R$ and $Q$ is a positive definite 
quadratic form. Actually, this can also be seen
without computation in the following way:  the
existence of the complex  constant $C_2$ and the
complex quadratic form $Q$  such that (\ref{A2})
holds is clear, and if $a$ is real  then
$Op^W(e^{-q_{\beta ,m}}\# a\# e^{-q_{\beta
,m}})=Op^W(e^{-q_{\beta ,m}})Op^W(a)Op^W(
e^{-q_{\beta ,m}})$ is a symmetric operator. As a
consequence $e^{-q_{\beta ,m}}\# a\# e^{-q_{\beta
,m}}$ must be real for $a$ real, which implies that
$Ce^{-Q}$ is real (and hence both $C$ and $Q$ are).
Moreover, one can show easily that the application 
$S_m(1)\ni a \mapsto e^{-q_{\beta ,m}}\# a\#
e^{-q_{\beta ,m}}$ maps continuously $S_m(1)$ into
${\cal S}(\Lambda_m )$, so that $Q$ is necessarily
positive definite.  \par When $a\equiv 1$, we get
from (\ref{2beta}), (\ref{A1}), (\ref{A2}) : 
\be
\label{A3} C_1C_2\int_{\Lambda_m}e^{-Q(x,\xi ,Y_1
)}dY_1=e^{-\tilde q_{2\beta ,m}}.
\ee Then the result follows  from (\ref{A1}),
(\ref{A2}), (\ref{A3}) by writing : $$ (e^{-q_{\beta
,m}}\# a\# e^{-q_{\beta ,m}})(\xxi )=\Big(
C_1C_2\int_{\Lambda_m} e^{-Q(x,\xi ,Y_1 )}dY_1\Big)
\int_{\Lambda_m}a(Y_1 )\left[e^{-Q(\xxi ,
Y_1)}dY_1\right]_N.$$ \par$\hfill\square$
\par\noindent  
{\bf 2. Proof of Lemma \ref{limit}}

Denote $\Delta (m)$  the difference between the two
expressions. For any $p,q\in\N$, we write:
\be
\Delta (m)=\Delta_1 (m,p,q) +\Delta_2 (m,p,q)
+\Delta_3 (m,p)+\Delta_4(p,q)
\ee 
with
\begin{eqnarray*}
 && 
\Delta_1 (m,p,q)= \\
&&
\quad \int_{\Lambda_m}
F(\Pi_{m_1}Q_{\beta ,q}\Pi_pX,
\Pi_{n}X)\,
\left[e^{-q_{\beta ,m} (X)}dX\right]_N
-\int_{\Lambda_\infty} F(\Pi_{m_1}Q_{\beta
,q}\Pi_pX,
\Pi_{n}X)\, d\hat\mu_\beta 
\end{eqnarray*}
\begin{eqnarray*}
\Delta_2 (m,p,q)=\int_{\Lambda_m}
\left( F(\Pi_{m_1}Q_{\beta ,m}\Pi_pX,
\Pi_{n}X) - F(\Pi_{m_1}Q_{\beta ,q}\Pi_pX,
\Pi_{n}X)\right) \,
\left[e^{-q_{\beta ,m} (X)}dX\right]_N \\ 
\Delta_3 (m,p)=\int_{\Lambda_m}
\left( F(\Pi_{m_1}Q_{\beta ,m}X,
\Pi_{n}X) - F(\Pi_{m_1}Q_{\beta ,m}\Pi_pX,
\Pi_{n}X)\right) \,
\left[e^{-q_{\beta ,m} (X)}dX\right]_N \\ 
\Delta_4(p,q)=
\int_{\Lambda_\infty} \left( F(\Pi_{m_1}Q_{\beta
,q}\Pi_pX,
\Pi_{n}X)-
F(\Pi_{m_1}Q_{\beta }X,
\Pi_{n}X)\right) \, d\hat\mu_\beta.
\end{eqnarray*} 
Now, by assumption on $F$, there
exists a positive constant
$C$ such that for any $m$ and $p$:
\be
\label{est1}
\vert \Delta_3 (m,p)\vert \leq C\int_{\Lambda_m}
\Vert \Pi_{m_1}Q_{\beta
,m}(X-\Pi_pX)\Vert_{\ell^\infty}
\left[e^{-q_{\beta ,m} (X)}dX\right]_N
\ee  and we have (with obvious notations):
\begin{eqnarray*}
\Vert \Pi_{m_1}Q_{\beta
,m}(X-\Pi_pX)\Vert_{\ell^\infty} && 
\leq \sup_{\vert i\vert\leq m_1}
\sum_{\vert j\vert >p}
\vert (Q_{\beta ,m})_{i,j}X_j\vert\\ &&
\leq 
\sup_{\vert i\vert\leq m_1}
\left(
\sum_{\vert j\vert >p} j^2\vert (Q_{\beta
,m})_{i,j}\vert^2\right)^{1/2}
\left( 1+\sum_{j\not= 0}\frac{X_j^2}{j^2}
\right)^{1/2}
\end{eqnarray*} and therefore, using (H5):
\be
\label{est2}
\Vert \Pi_{m_1}Q_{\beta
,m}(X-\Pi_pX)\Vert_{\ell^\infty}
\leq
\frac{C'}p\left( 1+\sum_{j\not= 0}\frac{X_j^2}{j^2}
\right).
\ee with a constant $C'$ independant of $m$ and $p$.
Since also
\be
\int_{\Lambda_m} X_j^2
\left[e^{-q_{\beta ,m} (X)}dX\right]_N =(Q_{\beta
,m}^{-1})_{j,j}\leq \Vert Q_{\beta
,m}^{-1}\Vert_{{\cal L}(\ell^2)}={\cal O}(1)
\ee uniformly with respect to $m$, we deduce from
(\ref{est1}), (\ref{est2}):
\be
\vert \Delta_3 (m,p)\vert ={\cal O}(p^{-1})
\ee uniformly with respect to $m$ and $p$.

In a similar way, using the fact that for fixed
$p$, both finite dimensional matrices
$\Pi_{m_1}Q_{\beta ,m}\Pi_p$ and $\Pi_{m_1}Q_{\beta
,q}\Pi_p$ tend to $\Pi_{m_1}Q_{\beta}\Pi_p$ as $m$
and $q$ tend to infinity, one can prove that:
\be
\Delta_2 (m,p,q)\rightarrow 0 \quad {\rm as}\quad
m\,\, {\rm and}\,\, q\rightarrow\infty .
\ee 
Moreover, for any fixed $(p,q)$ we see that
\be
\Delta_1 (m,p,q)\rightarrow 0 \quad {\rm as}\quad
m\rightarrow\infty .
\ee 
The same arguments also give, subtituing $Q_\beta$
to $Q_{\beta ,m}$, that
\be
\Delta_4 (p,q)\rightarrow 0 \quad {\rm as}\quad
p\,\, {\rm and}\,\, q\rightarrow\infty .
\ee
 Then, choosing $\varepsilon
>0$ arbitrarily small, one can first fix $p$ large
enough so that
$\vert \Delta_3 (m,p)\vert\leq\varepsilon $ for all
$(m,q)$ and $\vert\Delta_4
(p,q)\vert\leq\varepsilon $ for all $q$
sufficiently large, then fix
$q$ large enough so that $\vert \Delta_2 (m,p,
q)\vert
\leq\varepsilon $ for all sufficiently large $m$,
and finally get $\vert \Delta_1 (m,p, q)\vert
\leq\varepsilon $, and thus $\vert \Delta (m)\vert
\leq 4\varepsilon $, by taking $m$ large enough.
\hfill$\square$
\vskip 0.5cm
\par\noindent  {\bf 3. Proof of Lemma \ref{lemma3}}

\par From (\ref{8}) we get easily:
\begin{eqnarray} e^{-(\la F\xi ,\xi\ra +\la
Gx,x\ra)}\#  e^{-(\la F\xi ,\xi\ra +\la Gx,x\ra)}
&& =\pi^{-2(2m+1)}
\left\vert\int_{\Lambda_m}e^{2i\zeta y-\la G(x+y),
x+y\ra -\la F(\xi +\zeta ),\xi +\zeta\ra}
dyd\zeta\right\vert^2 \nonumber \\
\label{A21} && =\pi^{-2(2m+1)}\vert I\vert^2 
\end{eqnarray}
where
\be I = e^{-\la Gx,x\ra -\la F\xi ,\xi\ra}
\int e^{iy(2\zeta +2iGx)-\la Gy, y\ra -2\la F\xi
,\zeta\ra -\la F\zeta ,\zeta\ra}dyd\zeta .
\ee
Making the change of variables 
$\ds y'={\sqrt{2}}G^{\frac12}y$ and integrating
first with respect to $y'$ we get:
\be I = Ce^{-\la Gx,x\ra -\la F\xi ,\xi\ra}\int 
e^{-\la G^{-1}(\zeta +iGx),\zeta +iGx\ra -2\la F\xi
,\zeta\ra -\la F\zeta ,\zeta\ra}d\zeta
\ee
where $C$ is a constant, and therefore, setting 
$\ds\zeta'={\sqrt{2}}(F+G^{-1})^{\frac12}\zeta$,
this gives:
\be
\label{A22} I = C'e^{-\la F\xi ,\xi\ra-\la
(F+G^{-1})^{-1} (x-iF\xi ),x-iF\xi\ra}
\ee
where $C'$ is a constant. Then (\ref{A21})  and
(\ref{A22}) yield the result.
\hfill$\square$
\vskip 1.5cm\noindent
\section{Appendix 2: Remarks on Quantum Mixing
and Ergodicity}   
\setcounter{equation}{0}%
\setcounter{theorem}{0}%
\setcounter{proposition}{0}%
\setcounter{lemma}{0}%
\setcounter{corollary}{0}%
\setcounter{definition}{0}%
The elementary
remarks collected here, useful to clarify the
subsequent statements, are presumably 
known but
we were unable to locate a precise
reference. \par
We first formulate into an abstract
setting the definitions recalled in the
introduction. Let $H$ be a positive
self-adjoint operator on a separable Hilbert
space $\cal H$ such that $\sigma(H)$ is
discrete and simple and  $e^{-\beta H}$ is
trace-class for any positive $\beta$. Given
$A\in{\cal L}({\cal H})$ let $\omega(A)$ be
the quantum microcanonical measure defined
as in (\ref{qmc}) or the corresponding
quantum Gibbs measure at inverse temperature
$\beta$
\be
\label{Gibbs}
\omega(A) =\frac{\Tr\,(Ae^{-\beta H})}
{\Tr\,(e^{-\beta H})}
\ee
indifferently. Let also ${\cal A}$ be a
weakly closed sub-algebra of ${\cal
L}({\cal H})$ invariant under the action of
$e^{itH}$. In this general context we  assume
(and verify below in our specific case) the
existence of a family of normalized states
$(\psi_\lambda)_{\lambda\in\Lambda}$ complete
for
$\omega$ on ${\cal A}$, in the sense that
there is a probability measure
$d\nu(\lambda)$ on the set $\Lambda$ such
that  
\be
\label{traccia} 
\omega(A)=\int_{\Lambda}\la
A\psi_{\lambda},\psi_{\lambda}\ra_{\cal
H}\,d\nu(\lambda), \qquad
\forall\;A\in{\cal A}.
\ee
Then the quantum mixing property on ${\cal
A}$, {\it defined} as 
\be
\label{qmix}
\omega (A_H(t)B)\rightarrow 
\omega (A)\omega (B)\quad {\rm as}
\quad \vert t\vert \rightarrow \infty 
\ee 
for any operators
$A,B\in{\cal A}$
can be rewritten as 
\be
\label{qqmix}
\int_\Lambda \la A_H(t)B\psi_\lambda ,
\psi_\lambda\ra_{\cal H}d\nu (\lambda )
\rightarrow \int_\Lambda \la A\psi_\lambda ,
\psi_\lambda\ra_{\cal H}d\nu (\lambda )
\int_\Lambda \la B\psi_\lambda ,
\psi_\lambda\ra_{\cal H}d\nu (\lambda ).
\ee 
Remark that
(\ref{qmc}, \ref{Gibbs}, \ref{traccia})
imply the invariance property
\be
\label{inv} 
\int_{\Lambda}\la
A_H(t)\psi_{\lambda},\psi_{\lambda}\ra_{\cal
H}\,d\nu(\lambda)=\int_{\Lambda}\la
A\psi_{\lambda},\psi_{\lambda}\ra_{\cal
H}\,d\nu(\lambda)
\ee
and by analogy with the classical dynamical
systems, a natural possible definition of
quantum ergodicity is that for any $A\in{\cal
A}$:
\begin{equation}
\label{qerg'}
\frac1{T}\int_0^T\la
A_H(t)\psi_{\lambda},\psi_{\lambda}\ra_{\cal
H}\, dt\rightarrow \omega (A) \quad
\nu(\lambda)-{\rm a.e.}\quad {\rm as}\quad
{T\to\infty}.
\ee
This definition is also motivated from the
fact that in most situations the quantum
mixing property (\ref{qqmix}) implies
(\ref{qerg'}). Indeed, still assuming that
$\sigma (H)$ is discrete and simple, we see
in (\ref{qtimeaverage}) that for any $A\in
{\cal A}$ the limit
\be
\label{ta}
\lim_{T\to\infty}
\frac1{T}\int_0^T\la
A_H(t)\varphi,\psi\ra_{\cal
H}\, dt\,:=\la \bar A\varphi ,
\psi\ra_{\cal H}
\ee
exists for all $\varphi ,\psi\in {\cal H}$,
and the operator $\bar A\in {\cal A}$ is
invariant under the action of $e^{itH}$. As
a consequence, 
the
definition (\ref{qerg'}) of quantum
ergodicity is equivalent to the fact that
for any $A\in{\cal A}$ we have the identity
\be
\label{qerg"}
\la \bar A\psi_{\lambda} ,
\psi_{\lambda}\ra =\omega (A)
\ee
for $\nu
$-almost every $\lambda$. Now it is easy
to see that the quantum mixing property
implies that for any $A,B\in {\cal A}$,
$\omega (\bar AB)=\omega (A)\omega (B)$.
Therefore, if we assume moreover (which
is obviously true in the concrete example
discussed  below) that for any
$A\in{\cal A}$, the family
$(\om (AB))_{B\in{\cal A}}$ determines 
$\la  A\psi_{\lambda} ,
\psi_{\lambda}\ra $ for $\nu
$-almost every $\lambda$, we deduce
immediately that the quantum mixing implies
(\ref{qerg"}), and hence quantum ergodicity.
\vskip 0.2cm\noindent
Concerning this definition of quantum
ergodicity, we remark
that
it is trivially included in the notion of
ergodicity of the $W^*$ dynamical systems
\cite{Be,BR} with respect to the triple
$({\cal A}, \Theta, \phi)$ where $\Theta$
is the automorphism of ${\cal A}$ generated by
the unitary group $\ds e^{iHt}$ and $\phi$ is
the state defined by the microcanonical or
canonical measure. 
\vskip 0.2cm\noindent
Let us now turn to an explicit construction, 
in the
particular case
${\cal H}=L^2(\R^m)$ ($m<+\infty$ fixed)
mentioned in the introduction, of the
measures
$d\nu(\lambda)$, both in the microcanonical case
and in the canonical one as well, through some
natural choice of the set
$\{\psi_\lambda: \lambda\in\Lambda\}$.  This  
will also enable us to recover the classical 
definitions of mixing and ergodicity out of
(\ref{qmix}) and (\ref{qerg'}) at the
classical limit
$h\rightarrow 0$. 

More
precisely, for $\lambda =
(\lambda_x,\lambda_\xi )\in\R^{2m}$ consider
the Bargmann coherent states defined on
$\R^{2m}$:
\be
\label{csbis}
f_\lambda (x)=(\pi h)^{-m/4}
e^{ix\lambda_\xi/h - (x-\lambda_x)^2/2h} .
\ee
Then it is well known (see e.g.\cite{BS} Chapt.5)
that for any trace class operator $A$ on 
$L^2(\R^m )$, one has: 
\be 
\label{3'}
\int_{\R^{2m}}\la Af_\lambda ,
f_\lambda\ra_{L^2(\R^m)}d\lambda =\Tr\,(A)\; .
\ee 
In particular, since 
$e^{-\beta H}$ is trace class on $L^2(\R^m
)$, then 
$$ 
\int \Vert e^{-\beta
H/2}f_\lambda \Vert^2 d\lambda  = \int \la
e^{-\beta H}f_\lambda ,f_\lambda \ra
d\lambda = \Tr\,(e^{-\beta H})<+\infty 
$$ 
Hence we can consider the following
probability measures on $\R^{2m}$: 
\be 
d\nu_m(\lambda )=\frac{\Vert e^{-\beta
H/2}f_\lambda \Vert^2 d\lambda}{\int \Vert 
e^{-\beta H/2}f_\lambda \Vert^2 d\lambda},
\quad  
d\nu_{\Delta,E}(\lambda )=\frac{\Vert
{\delta(H-E)}
f_\lambda \Vert^2
d\lambda}{\int \Vert
{\delta(H-E)}f_\lambda
\Vert^2 d\lambda}.
\ee 
where as in \S 1,  $\ds
\delta(H-E)=\sum_{n:E-\Delta<E_n<E}P_n$
with $\Delta >0$ fixed.  If we also set 
\be
 \psi^c_\lambda =
\frac{e^{-\beta H/2}f_\lambda}{\Vert
e^{-\beta H/2}f_\lambda \Vert} 
\, ,\quad
 \psi^{mc}_\lambda=
\frac{
{\delta(H-E)}
f_\lambda}{ \Vert
{\delta(H-E)}f_\lambda
\Vert }.
\ee 
then we
have the following result (to be compared
with (\ref{traccia})):
\begin{lemma} 
For any
bounded operator $A$ on  $L^2(\R^m )$, the
following identities hold: 
$$
\frac{\Tr\,\left( Ae^{-\beta H}\right)}{{\rm
Tr}\left( e^{-\beta H}\right)}=\int \la
A\psi^c_\lambda ,
\psi^c_\lambda\ra d\nu_m (\lambda )
,\quad
\frac{\Tr\,\left(
A\delta(H-E)\right)}{{\rm Tr}\left(
\delta(H-E)\right)}=\int \la
A\psi^{mc}_\lambda ,
\psi^{mc}_\lambda\ra d\nu_{\Delta,E} (\lambda ).
$$
\end{lemma} 
{\it Proof}- Just write:
\begin{eqnarray*}
&& \Tr\,\left( Ae^{-\beta H}\right)={\rm
Tr}\left( e^{-\beta H/2} Ae^{-\beta
H/2}\right),
\\
&&
\Tr\,\left(
A\delta(H-E)\right)={\rm Tr}(
{\delta(H-E)} A
{\delta(H-E)})
\end{eqnarray*}
and use
(\ref{3'}). 
\hfill$\square$.\par
Consider now the particular
case where $A$ is the $h$-Weyl quantization of a
classical observable
$a=a(x,\xi )\in {\cal S} (\R^{2m})$, namely the
operator $Op_h^W(a)$ defined by the oscillatory
integral:
\be
\label{scWeyl} Op_h^W(a)u(x) =(2\pi h)^{-m}\int
e^{i (x-y)\xi
/h}a\left(\frac{x+y}{2},\xi\right)u(y)\,dy\,d\xi .
\ee 
A well known direct application of the stationary
phase method yields
$$
\lim_{h\rightarrow 0}\la Op_h^W(a)f_\lambda
,f_\lambda \ra = a(\lambda ).
$$ 
As a
consequence, if we also have
$$H=Op_h^W(q)$$ for some symbol $q\in C^{\infty}
(\R^{2m})$, then the semiclassical symbolic and
functional calculus of pseudodifferential
operators (see \cite{Ro}) immediately implies:
\begin{lemma}
\label{lemma2'} 
For any
$\lambda\in\R^{2m}$,
$$
\lim_{h\rightarrow 0}\la Op_h^W(a) \psi_\lambda
,\psi_\lambda \ra = a(\lambda ) .
$$ 
Moreover,
\begin{eqnarray*}
&&
\lim_{h\rightarrow 0}\int_{\R^{2m}}\la
Op_h^W(a)\psi^c_\lambda ,\psi^c_\lambda \ra 
d\nu_m (\lambda ) =
\int a(\lambda )d\mu^c (\lambda )
 \\
&&
d\mu^c (\lambda )=
\frac{e^{-\beta q}d\lambda}{\int_{\R^{2m}}
e^{-\beta q}d\lambda}
\end{eqnarray*}
and
\begin{eqnarray*} 
&& 
\lim_{h\rightarrow 0}\int_{\R^{2m}}\la
Op_h^W(a)\psi^{mc}_\lambda ,\psi^{mc}_\lambda \ra 
d\nu_{\Delta,E} (\lambda ) =
\int_{\R^{2m}} a(\lambda )d\mu^{mc} (\lambda )
\\
&&
\quad  d\mu^{mc} (\lambda )=
\frac{\delta{(q-E)}d\lambda}{\int_{\R^{2m}}
\delta{(q-E)}d\lambda}
\end{eqnarray*}
\end{lemma} 
Since, by the semiclassical Egorov theorem
(\cite{Ro}, \S 5.4) the principal symbol of
\linebreak
$\ds e^{itH/h}Op_h^W(a)e^{-itH/h}$ is given by
$$
a_t(x,\xi )=a\big( \phi_t(x,\xi )\big)
$$
where
$\phi_t$ is the Hamiltonian flow generated by $q$,
it follows that for any $a,b\in {\cal
S}(\R^{2m})$:
\be\label
{scmix}
\lim_{h\rightarrow 0}\la
e^{itH/h}Op_h^W(a)e^{-itH/h}Op_h^W(b)
\psi_\lambda
 ,\psi_\lambda \ra =
a\big(\phi_t(\lambda )
\big) b(\lambda ).
\ee 
It has now become clear out of Lemma
\ref{lemma2'} and (\ref{scmix}) that the quantum
notions of mixing and ergodicity given by
(\ref{qmix}) and (\ref{qerg'}) formally yield
the corresponding classical notions as
$h\rightarrow 0$.

As a final remark let us mention that if $A$
is a pseudodifferential operator also the Von
Neumann definition (\ref{qtimeaverage}) reproduces
the classical one at the classical limit if
$\la u_n,A u_n\ra$ tends to the phase average of
the symbol of $A$, as verified in many instances (see
e.g.\cite{Sc,CdV,HMR,Ze1,Ze2,DEGI}), in which $H$ is
the  quantization of a \ha\ generating an ergodic
flow. Some authors (\cite{Sa,Ze2} assume this
limiting property as the very definition of quantum
ergodicity.

\end{document}